\documentclass{llncs}

\institute{CNRS (LaBRI)}
\author{Richard Moot}
\title{Proof nets for the Displacement calculus}

\usepackage{tikz}
\usetikzlibrary{arrows,matrix,graphs,shapes,snakes,automata,backgrounds,petri,positioning,calc,decorations.pathmorphing}
\usepackage[leqno]{amsmath}
\usepackage{amssymb}
\usepackage{latexsym}
\usepackage{stmaryrd}
\usepackage{harvard}
\usepackage{proof}

\newcommand{\editout}[1]{}
\newcommand{\splitc}{\,\check{\,}}
\newcommand{\bridge}{\,\hat{\,}}

\newcommand{\apsarrow}{\mapsto}

\newcommand{\downl}{\downarrow_{>}}
\newcommand{\downr}{\downarrow_{<}}
\newcommand{\downk}{\downarrow_{k}}
\newcommand{\upl}{\uparrow_{>}}
\newcommand{\upr}{\uparrow_{<}}
\newcommand{\upk}{\uparrow_{k}}
\newcommand{\prodl}{\odot_{>}}
\newcommand{\prodr}{\odot_{<}}
\newcommand{\prodk}{\odot_{k}}
\newcommand{\ldl}{\mathbin{\backslash}}
\newcommand{\ldr}{\mathbin{/}}
\newcommand{\lpr}{\mathbin{\bullet}}
\newcommand{\timesk}{\mathbin{\times_k}}
\newcommand{\timesl}{\mathbin{\times_{>}}}
\newcommand{\timesr}{\mathbin{\times_{<}}}
\newcommand{\one}{\mathbf{1}}
\newcommand{\dplus}{\!+\!}

\tikzset{pas/.style={fill=gray!60}, 
act/.style={fill=gray!30},
main/.style={draw,fill=white},
ctx/.style={rounded rectangle,minimum size=7mm},
val/.style={rectangle,minimum size=7mm},
cmd/.style={chamfered rectangle,draw,fill=white},
tns/.style={circle,minimum size=5mm,draw,fill=white},
par/.style={circle,minimum size=5mm,draw,fill=black}, 
minipar/.style={circle,minimum size=2.5mm,draw,fill=black}, 
pn/.style={rounded corners, rectangle,fill=blue!30,draw,minimum size=15mm},
medpn/.style={rounded corners, rectangle,fill=blue!30,draw,minimum size=20mm},
 bigpn/.style={rounded corners, rectangle,fill=blue!30,draw,minimum size=25mm}}

\begin{document}
\maketitle

\section{Introduction}

The Displacement calculus was introduced by \citeasnoun{mvf11displacement} as an extension of the Lambek calculus with discontinuous operators. These discontinuous
connectives  allow the Displacement calculus to solve
a large number of problems with the Lambek calculus. Examples of the phenomena treated by \citeasnoun{mvf11displacement} include discontinuous idioms (such as ``ring up''
and ``give the cold shoulder''), quantifier scope, extraction (including pied-piping) and gapping.

This paper extends earlier work by 
\citeasnoun{mf08disco}, \citeasnoun{moot13lambek} and \citeasnoun{ov14}, combining the strengths
of these different approaches while at the same time diminishing the inconveniences. Notably, it is the first proof net calculus which does not operate by translation
into some other logic, but provides proof nets for the Displacement calculus directly.

\section{The Displacement calculus}

The presentation of the Displacement calculus
closely follows the natural deduction calculus used by \citeasnoun{mvf11displacement}. String
terms are built over a countably infinite alphabet of variables (for
readability, we will often used natural language words as if they were variables), a
special separator symbol ``$\one$'', where string concatenation is denoted by
``+'' (a binary, associative infix operator on string terms). As
usual, $\epsilon$ denotes the empty string. The
\emph{sort} of a string term is the number of occurrences of the
separator ``$\one$''.

I  use lower-case roman
letters $p$, $q$ $\ldots$ for atomic string terms (for enhanced
readability, I will often use the standard convention of using words from
the lexicon in the place of such atomic string terms),  lower-case roman
letters $a$, $b$, $\ldots$ for string terms without separator symbols
and lower-case greek letters $\alpha$, $\beta$, $\ldots$ for strings
containing any number of separator symbols. So the string term $p +
\one + q + \one + r$ is a string of sort 2 with three atomic subterms.

The key to the Displacement calculus is the wrap operator
$\alpha \timesk \beta$. There is some minor variation in the definition of this
operator: sometimes \cite{mvf11displacement} $k$ is either the constant ``$>$'' or the constant
``$<$'' (in which case $\alpha$ is of sort greater than zero and the
denotation of the term replaces respectively the first and the last
occurrences of $\one$ in $\alpha$ by $\beta$. Sometimes \cite{morrill2010} $k$ is an
integer (between 1 and the sort of $\alpha$) and $\alpha \timesk
\beta$ replaces the $k$th separator in $\alpha$ by
$\beta$. The equations below given the definition of ``$\timesk$''.
\begin{align}
\label{def:wleft}(a + \one + \alpha) \timesl \beta &=_{\mathit{def}} a + \beta + \alpha \\
\label{def:wright} (\alpha + \one + a) \timesr \beta &=_{\mathit{def}} \alpha + \beta + a \\
(a_1 + \one + \ldots + a_n + \one +\alpha)\times_n \beta &=_{\mathit{def}}  a_1 + \one + \ldots + a_n + \beta +\alpha
\end{align}

Where the Lambek calculus connectives get their meaning with respect
to concatenation ``+'', the discontinuous connectives of the
Displacement calculus get their meaning with respect to ``$\timesk$''
(this entails different connectives for different values of $k$). The standard interpretation of the Lambek calculus connectives
for string models, with ``+'' denoting concatenation, is the following.
\begin{align}
| A\mathbin{\backslash} C | &=_{\mathit{def}} \{ \beta \,|\, \forall \alpha \in | A |,
\alpha+\beta \in | C | \} \\
| C\mathbin{/} B | &=_{\mathit{def}} \{ \alpha \,|\, \forall \beta \in | B |,
\alpha+\beta \in | C | \} \\
| A \mathbin{\bullet} B | & =_{\mathit{def}} \{ \alpha+\beta \,|\, \alpha
\in | A | \,\wedge\, \beta \in | B | \}
\end{align}
The discontinuous connectives of the Displacement calculus use
``$\timesk$'' instead of ``+'' (we present only the connectives for
$>$ here). 
\begin{align}
| A\downl C | &=_{\mathit{def}} \{ \beta \,|\, \forall \alpha \in | A |,
\alpha\timesl\beta \in | C | \} \\
| C\upl B | &=_{\mathit{def}} \{ \alpha \,|\, \forall \beta \in | B |,
\alpha\timesl\beta \in | C | \} \\
| A \prodl B | & =_{\mathit{def}} \{ \alpha\timesl\beta \,|\, \alpha
\in | A | \,\wedge\, \beta \in | B | \}
\end{align}
We can further unfold these definitions, using Definition~\ref{def:wleft} for ``$\timesl$''  to obtain.
\begin{align}
\label{def:unfoldstart}| A\downl C | &=_{\mathit{def}} \{ \beta \,|\, \forall (a + \one + \alpha) \in | A |,
a + \beta + \alpha \in | C | \} \\
| C\upl B | &=_{\mathit{def}} \{ (a + \one + \alpha) \,|\, \forall \beta \in | B |,
a + \beta + \alpha \in | C | \} \\
\label{def:unfoldend}| A \prodl B | & =_{\mathit{def}} \{ a + \beta + \alpha \,|\, (a + \one + \alpha)
\in | A | \,\wedge\, \beta \in | B | \}
\end{align}
Given these definitions, the meaning of $A\downl C$ is defined as the
set of expressions which select a circumfix $A$, which wraps around
the string denoted by $A\downl C$ to form an expression
$C$. Similarly, $C\upl B$ extracts a $B$ formula not occurring after a separator.
\subsection{Formulas and sorts}

We have already defined the sort of a string term as the number of
occurrences of the separator constant ``$\one$''. The \emph{sort} of a formula corresponds to the number of separators ``$\one$'' occurring in its denotation. That is, a formula of sort $n$ is assigned a string term of the form $a_0 + \one + \ldots + \one + a_n$ (with all $a_i$
of sort 0 according to our notational convention).
For a given grammar, its signature defines the sort of all atomic
formulas occurring in the grammar.
We assume throughout that the atomic formulas
$\mathit{s}$, $\mathit{n}$, $\mathit{np}$, $\mathit{pp}$ have sort 0
(some other atomic formulas, such as $\mathit{inf}$ when used for Dutch verb clusters, have sort 1).

Table~\ref{tab:sorts} shows how to compute the sort of complex formulas. All subformulas of a formula are assigned a sort, so when we
compute $s(C\mathbin{/}B)$ using its entry in Table~\ref{tab:sorts} we know that $s(C) \geq s(B)$, because if not, then $s(C\mathbin{/}B)$ would be less than zero and therefore
not a valid (sub)formula (similar constraints can be derived from the other implications, eg.\ we can show that $s(C\mathbin{\uparrow}B) \geq 1$).

\begin{table}
\begin{align*}
s(A\mathbin{\bullet} B) &= s(A) + s(B) \qquad\qquad&s(A\mathbin{\odot} B) &= s(A) + s(B) -1 && \scriptstyle{s(A) \geq 1}\\
s(A\mathbin{\backslash} C) &= s(C) - s(A)  \qquad\qquad&s(A\mathbin{\downarrow} C) &= s(C) + 1 - s(A) && \scriptstyle{s(A) \geq 1}
\\
s(C\mathbin{/}B) &= s(C) - s(B)\qquad\qquad&  s(C\mathbin{\uparrow}B) &= s(C) + 1 - s(B) && \scriptstyle{s(C) \geq s(B)}\\
\end{align*}
\caption{Computing the sort of a complex formula given the sort of its
  immediate subformulas}
\label{tab:sorts}
\end{table}

As an example, following \citeasnoun{mvf11displacement}, we can assign a discontinuous lexical entry like ``give the cold shoulder'' the lexical formula
$(np\mathbin{\backslash} s)\upl np$ and string term $\textit{gave} + \one + \textit{the} + \textit{cold} + \textit{shoulder}$ (of the required sort 1).

\subsection{Natural deduction rules}

Figures~\ref{fig:ndlambek} and \ref{fig:ndlw} 
 give the
natural deduction rules for the Lambek calculus and for the left wrap rules respectively (the other wrap rules follow the same pattern). The left wrap rules of Figure~\ref{fig:ndlw} correspond rather closely to
the interpretation of the formulas given in
Definitions~\ref{def:unfoldstart} to~\ref{def:unfoldend}.

\begin{figure}

$$
\begin{array}{ccc}
\infer[\backslash E]{\alpha\dplus\gamma:C}{\alpha:A & \gamma:A\mathbin{\backslash} C} &&
\infer[\backslash I_i]{\gamma:A\mathbin{\backslash} C}{\infer*{\alpha\dplus\gamma:C}{[ \alpha:A ]^i}}\\
\\
\infer[/ E]{\gamma\dplus\beta:C}{\gamma:C\mathbin{/} B & \beta:B} &&
\infer[/ I_i]{\gamma:C\mathbin{/} B}{\infer*{\gamma\dplus\beta:C}{[ \beta:B ]^i}}
\\
\\
\infer[\bullet E_i]{\gamma[\delta]:C}{\delta:A\bullet B & \infer*{\gamma[\alpha\dplus\beta]:C}{[\alpha:A]^i & [\beta:B]^i}}
&&
\infer[\bullet I]{\alpha\dplus\beta:A\bullet B}{\alpha:A & \beta:B}
\end{array}
$$

\caption{Proof rules -- Lambek calculus}
\label{fig:ndlambek}
\end{figure}

\begin{figure}

$$
\begin{array}{ccc}
\infer[\downl E]{a\dplus\gamma\dplus\alpha:C}{a\dplus\one\dplus\alpha:A & \gamma:A\downl C} &&
\infer[\downl I_i]{\gamma:A\downl C}{\infer*{a\dplus\gamma\dplus\alpha:C}{[a\dplus\one\dplus\alpha:A]^i}}
\\
\\
\infer[\upl E]{c\dplus\beta\dplus\gamma:C}{c\dplus\one\dplus\gamma:C\upl B & \beta:B} &&
\infer[\upl I_i]{c\dplus\one\dplus\gamma:C\upl B}{\infer*{c\dplus\beta\dplus\gamma:C}{[\beta:B]^i}}
\\
\\
\infer[\prodl E_i]{\gamma[\delta]:C}{\delta:A\prodl B & \infer*{\gamma[a\dplus\beta\dplus\alpha]:C}{[a\dplus\one\dplus\alpha:A]^i & [\beta:B]^i}}
&&
\infer[\prodl I]{a\dplus\beta\dplus\alpha:A\prodl B}{a\dplus\one\dplus\alpha:A & \beta:B}
\end{array}
$$

\caption{Proof rules --- leftmost infixation,extraction}
\label{fig:ndlw}
\end{figure}

\editout{
\begin{figure}

$$
\begin{array}{ccc}
\infer[\downr E]{\alpha\dplus\gamma\dplus a:C}{\alpha\dplus\one\dplus a:A & \gamma:A\downr C} &&
\infer[\downr I_i]{\gamma:A\downr C}{\infer*{\alpha\dplus\gamma\dplus a:C}{[\alpha\dplus\one\dplus a:A]^i}}\\
\\
\infer[\upr E]{\gamma\dplus\beta\dplus c:C}{\gamma\dplus\one\dplus c:C\upr B & \beta:B} &&
\infer[\upr I_i]{\gamma\dplus\one\dplus c:C\upr B}{\infer*{\gamma\dplus\beta\dplus c:C}{[\beta:B]^i}}
\\
\\
\infer[\prodr E_i]{\gamma[\delta]:C}{\delta:A\prodr B & \infer*{\gamma[\alpha\dplus\beta\dplus a]:C}{[\alpha\dplus\one\dplus a:A]^i & [\beta:B]^i}}
&&
\infer[\prodr I]{\alpha\dplus\beta\dplus a:A\prodr B}{\alpha\dplus\one\dplus a:A & \beta:B}
\end{array}
$$

\caption{Proof rules --- rightmost infixation,extraction}
\label{fig:ndrw}
\end{figure}
}

\section{Proof nets}

One of the goals of proof search in type-logical grammars is to
enumerate all possible readings for a given sentence.
The bureaucratic aspects of the sequent calculus proof search make it
hard to use sequent calculus directly for this goal, since
sequent calculus allows a great number of inessential rule
permutations. The situation for natural deduction is somewhat better, 
since the proof rules correspond directly to steps in meaning composition,
even though there is still a large number of possible rule permutations for the $\bullet E$ and
$\prodk E$ rules.

Proof nets are a way of representing proofs which removes the
``bureaucratic'' aspects of sequent proofs and simplifies the product
rules of Lambek calculus natural deduction.
One of the open questions of \citeasnoun{morrill2010} is whether the Displacement calculus has a proof net calculus. 

\citeasnoun{ov14} provides a translation of the Displacement calculus to a multimodal system. However, this system uses a rather large set of structural rules and these rules are defined
modulo equivalence classes, which makes their use in existing multimodal theorem provers \cite{moot07filter} difficult. In this section, I will extend the proof net calculus for the
Lambek calculus of \citeasnoun{mp} to the Displacement calculus. I will, in particular, provide an efficiently checkable correctness condition in the form of graph contractions.

\subsection{Links}

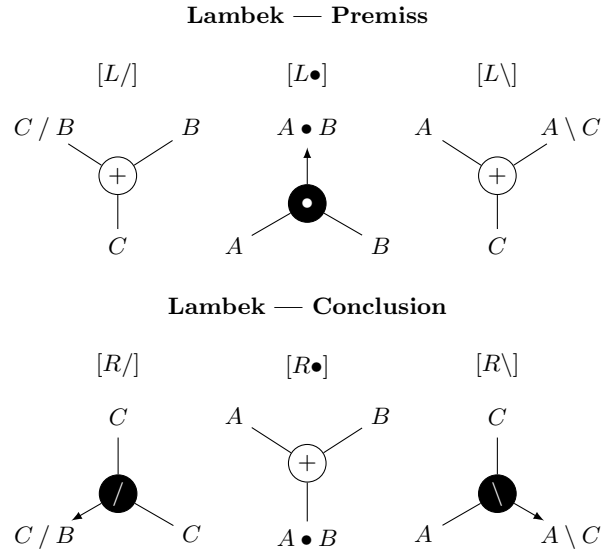
\begin{figure}
\begin{center}
\begin{tabular}{ccc}
\multicolumn{3}{c}{\textbf{Lambek --- Premiss}} \\[3mm]
\begin{tikzpicture}
\node (lab) at (3em,7em) {$[L \ldr]$};
\node (ab) at (3em,0.0em) {$C$};
\node (a) at (0,4.8em) {$\smash{C\ldr B}\rule{0pt}{1.3ex}$};
\node (b) at (6em,4.8em) {$B_{\rule{0pt}{1.2ex}}$};
\node[tns] (c) at (3em,2.868em) {};
\node (clab) at (3em,2.868em) {+};
\draw (c) -- (ab);
\draw (c) -- (a);
\draw (c) -- (b);
\end{tikzpicture} &
\begin{tikzpicture}
\node (lab) at (3em,7em) {$[L \lpr]$};
\node (ab) at (3em,4.8em) {$A\lpr B_{\rule{0pt}{1.2ex}}$};
\node (a) at (0,0) {$A$};
\node (b) at (6em,0) {$B$};
\node[par] (c) at (3em,1.732em) {};
\node (clab) at (3em,1.732em) {\textcolor{white}{$\bullet$}};
\path[>=latex,->] (c) edge (ab);
\draw (c) -- (a);
\draw (c) -- (b);
\end{tikzpicture} &
\begin{tikzpicture}
\node (lab) at (3em,7em) {$[L \ldl]$};
\node (ab) at (3em,0.0em) {$C$};
\node (a) at (0,4.8em) {$A_{\rule{0pt}{1.2ex}}$};
\node (b) at (6em,4.8em) {$\smash{A\ldl C}\rule{0pt}{1.3ex}$};
\node[tns] (c) at (3em,2.868em) {};
\node (clab) at (3em,2.868em) {+};
\draw (c) -- (ab);
\draw (c) -- (a);
\draw (c) -- (b);
\end{tikzpicture} \\[3mm]
\multicolumn{3}{c}{\textbf{Lambek --- Conclusion}} \\[3mm]
\begin{tikzpicture}
\node (lab) at (3em,7em) {$[R \ldr]$};
\node (ab) at (3em,4.8em) {$C_{\rule{0pt}{1.2ex}}$};
\node (a) at (0,0) {$\smash{C\ldr B}\rule{0pt}{1.3ex}$};
\node (b) at (6em,0) {$C_{\rule{0pt}{1.2ex}}$};
\node[par] (c) at (3em,1.732em) {};
\node (clab) at (3em,1.732em) {\textcolor{white}{$/$}};
\draw (c) -- (ab);
\path[>=latex,->] (c) edge (a);
\draw (c) -- (b);
\end{tikzpicture} &
\begin{tikzpicture}
\node (lab) at (3em,7em) {$[R \lpr]$};
\node (ab) at (3em,0.0em) {$A\lpr B$};
\node (a) at (0,5.0em) {$A$};
\node (b) at (6em,5.0em) {$B$};
\node[tns] (c) at (3em,3.068em) {};
\node (clab) at (3em,3.068em) {+};
\draw (c) -- (ab);
\draw (c) -- (a);
\draw (c) -- (b);
\end{tikzpicture} &
\begin{tikzpicture}
\node (lab) at (3em,7em) {$[R \ldl]$};
\node (ab) at (3em,4.8em) {$C_{\rule{0pt}{1.2ex}}$};
\node (a) at (0,0) {$A_{\rule{0pt}{1.2ex}}$};
\node (b) at (6em,0) {$\smash{A\ldl C}\rule{0pt}{1.3ex}$};
\node[par] (c) at (3em,1.732em) {};
\node (clab) at (3em,1.732em) {\textcolor{white}{$\backslash$}};
\draw (c) -- (ab);
\draw (c) -- (a);
\path[>=latex,->] (c) edge (b);
\end{tikzpicture} \\[3mm]
\end{tabular}
\end{center}
\caption{Links for the Lambek calculus connectives of the Displacement
calculus}
\label{fig:lambeklinks}
\end{figure}

\begin{figure}
\begin{center}
\begin{tabular}{ccc}
\multicolumn{3}{c}{\textbf{Discontinuous --- Premiss}} \\[3mm]
\begin{tikzpicture}
\node (lab) at (3em,7em) {$[L \upk]$};
\node (ab) at (3em,0.0em) {$C$};
\node (a) at (0,4.8em) {$\smash{C\upk B}\rule{0pt}{1.3ex}$};
\node (b) at (6em,4.8em) {$B_{\rule{0pt}{1.2ex}}$};
\node[tns] (c) at (3em,2.868em) {};
\node (clab) at (3em,2.868em) {$\timesk$};
\draw (c) -- (ab);
\draw (c) -- (a);
\draw (c) -- (b);
\end{tikzpicture} &
\begin{tikzpicture}
\node (lab) at (3em,7em) {$[L \prodk]$};
\node (ab) at (3em,4.8em) {$A\prodk B_{\rule{0pt}{1.2ex}}$};
\node (a) at (0,0) {$A$};
\node (b) at (6em,0) {$B$};
\node[par] (c) at (3em,1.732em) {};
\node (clab) at (3em,1.732em) {\textcolor{white}{$\prodk$}};
\path[>=latex,->] (c) edge (ab);
\draw (c) -- (a);
\draw (c) -- (b);
\end{tikzpicture} &
\begin{tikzpicture}
\node (lab) at (3em,7em) {$[L \downk]$};
\node (ab) at (3em,0.0em) {$C$};
\node (a) at (0,4.8em) {$A_{\rule{0pt}{1.2ex}}$};
\node (b) at (6em,4.8em) {$\smash{A\downk C}\rule{0pt}{1.3ex}$};
\node[tns] (c) at (3em,2.868em) {};
\node (clab) at (3em,2.868em) {$\timesk$};
\draw (c) -- (ab);
\draw (c) -- (a);
\draw (c) -- (b);
\end{tikzpicture} \\[3mm]
\multicolumn{3}{c}{\textbf{Discontinuous --- Conclusion}} \\[3mm]
\begin{tikzpicture}
\node (lab) at (3em,7em) {$[R \upk]$};
\node (ab) at (3em,4.8em) {$C_{\rule{0pt}{1.2ex}}$};
\node (a) at (0,0) {$\smash{C\upk B}\rule{0pt}{1.3ex}$};
\node (b) at (6em,0) {$B_{\rule{0pt}{1.2ex}}$};
\node[par] (c) at (3em,1.732em) {};
\node (clab) at (3em,1.732em) {\textcolor{white}{$\upk$}};
\draw (c) -- (ab);
\path[>=latex,->] (c) edge (a);
\draw (c) -- (b);
\end{tikzpicture} &
\begin{tikzpicture}
\node (lab) at (3em,7em) {$[R \prodk]$};
\node (ab) at (3em,0.0em) {$A\prodk B$};
\node (a) at (0,5.0em) {$A$};
\node (b) at (6em,5.0em) {$B$};
\node[tns] (c) at (3em,3.068em) {};
\node (clab) at (3em,3.068em) {$\timesk$};
\draw (c) -- (ab);
\draw (c) -- (a);
\draw (c) -- (b);
\end{tikzpicture} &
\begin{tikzpicture}
\node (lab) at (3em,7em) {$[R \downk]$};
\node (ab) at (3em,4.8em) {$C_{\rule{0pt}{1.2ex}}$};
\node (a) at (0,0) {$A_{\rule{0pt}{1.2ex}}$};
\node (b) at (6em,0) {$\smash{A\downk C}\rule{0pt}{1.3ex}$};
\node[par] (c) at (3em,1.732em) {};
\node (clab) at (3em,1.732em) {\textcolor{white}{$\downk$}};
\draw (c) -- (ab);
\draw (c) -- (a);
\path[>=latex,->] (c) edge (b);
\end{tikzpicture} \\[3mm]
\end{tabular}
\end{center}
\caption{Links for the discontinuous connectives of the Displacement
calculus}
\label{fig:discolinks}
\end{figure}
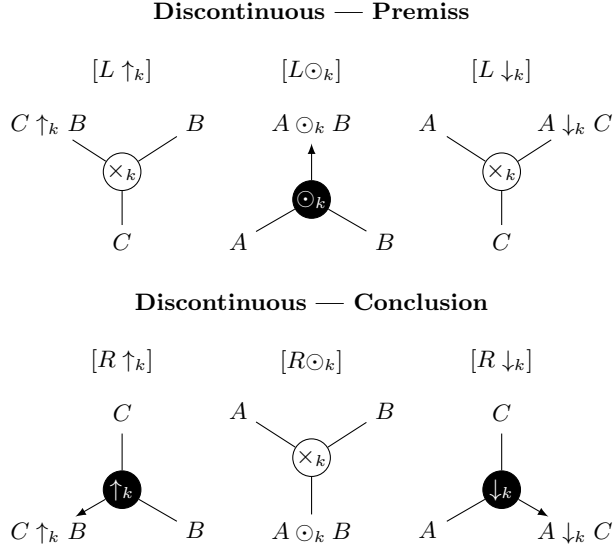

Figures~\ref{fig:lambeklinks} and \ref{fig:discolinks} show the links
for Displacement calculus proof structures. Each link connects three
formulas to a central node. The formulas written above
the central node of a link are the \emph{premisses} of the link, the
formulas written below it are its \emph{conclusions}. The linear order
of both the premisses and the conclusions of a link is important.

We distinguish between par links, where the central node is
filled black, and tensor links, where the central node is not filled
(this is the familiar tensor/par distinction of multiplicative linear logic).
Par nodes are further distinguished by an arrow pointing to the main formula
of the link.

\subsection{Proof structures}

A \emph{proof structure} is a set of formulas and a set of links such
that.
\begin{enumerate}
\item each link instantiates one of the links shown in
  Figures~\ref{fig:lambeklinks} and \ref{fig:discolinks} (for specific values of $A$, $B$, $C$ and $k$),
\item each formula is the premiss of at most one link,
\item each formula
is the conclusion of at most one link. 
\end{enumerate}

Formulas which are not the premiss of any link are the conclusions of
the proof structure. Formulas which are not the conclusion of any link
are the hypotheses of the proof structure (the word ``conclusion'' is overloaded: we talk about conclusions of proofs, conclusions of rules, conclusions of links and conclusions of proof
structures; when the intended use is clear from the context, I will often simply use the word ``conclusion'' without further qualification). The \emph{inputs} of a proof
structure are its hypotheses and the active conclusions of its par
links (that is, the conclusions of all par links in the proof structure except, for the implications, the one
with the arrow); we will call the inputs which are not hypotheses the
\emph{auxiliary inputs} of a proof structure.

To construct a proof structure for a given sequent $A_1,\ldots,A_n \vdash C$,
we unfold the $A_i$ as premisses and $C$ as a conclusion. This will provide a proof structure with (atomic) conclusions other than $C$ and (atomic) premisses other than the $A_i$.  We
identify these atomic hypotheses with atomic conclusions (of the same atomic formula) until we obtain a proof structure of $A_1,\ldots,A_n \vdash C$. This can fail if an
atomic formula has more occurrences as a hypothesis than as a conclusion (as it should, since such sequents are underivable). 

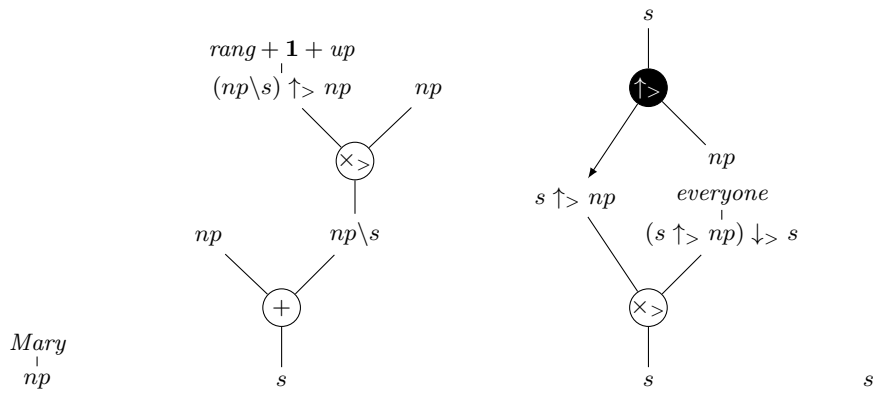
\begin{figure}[bt]
\begin{tikzpicture}
\node (mary) at (2em,0em) {$np$};
\node (mlab) at (2em,1.5em) {$\textit{Mary}$};
\draw (2em,0.6em) -- (2em,1.0em);
\node (goals) at (36em,0em) {$s$};
\node (wtv) at (12em,12em) {$(np\backslash s) \upl np$};
\node (rlab) at (12em,13.5em) {$\textit{rang}+\one+\textit{up}$};
\draw (12em,12.6em) -- (12em,13.0em);
\node (obj) at (18em,11.85em) {$np$};
\node[tns] (t1) at (15em,9em) {};
\node (t1lab) at (15em,9em) {$\timesl$};
\node (iv) at (15em,6em) {$np\backslash s$};
\node (subj) at (9em,5.85em) {$np$};
\node[tns] (t2) at (12em,3em) {};
\node (t2lab) at (12em,3em) {+};
\node (s1) at (12em,0em) {$s$};
\draw (t2) -- (s1);
\draw (t2) -- (subj);
\draw (t2) -- (iv);
\draw (t1) -- (iv);
\draw (t1) -- (wtv);
\draw (t1) -- (obj);
\node (gq) at (30em,6em) {$(s\upl np)\downl s$};
\node (ql) at (30em,7.5em) {$\textit{everyone}$};
\draw (30em,6.6em) -- (30em,7.0em);
\node (snp) at (24em,7.5em) {$s\upl np$};
\node (sg) at (27em,0em) {$s$};
\node[tns] (t3) at (27em,3em) {};
\node (t3lab) at (27em,3em) {$\timesl$};
\draw (t3) -- (gq);
\draw (t3) -- (snp);
\draw (t3) -- (sg);
\node (qnp) at (30em,9em) {$np$};
\node (sp) at (27em,15em) {$s$};
\node[par] (par) at (27em,12em) {};
\node (parlab) at (27em,12em) {\textcolor{white}{$\upl$}};
\draw[>=latex,->] (par) -- (snp);
\draw (par) -- (sp);
\draw (par) -- (qnp);
\end{tikzpicture}
\caption{Unfolding for the sentence ``Mary rang everyone up''.}
\label{fig:unfold}
\end{figure}

\begin{figure}[bt]
\begin{tikzpicture}
\node (mlab) at (11em,9.5em) {$\textit{Mary}$};
\draw (11em,8.6em) -- (11em,9.0em);
\node (wtv) at (14em,14em) {$(np\backslash s) \upl np$};
\node (rlab) at (14em,15.5em) {$\textit{rang}+\one+\textit{up}$};
\draw (14em,14.6em) -- (14em,15.0em);
\node (obj) at (20em,13.85em) {$np$};
\node[tns] (t1) at (17em,11em) {};
\node (t1lab) at (17em,11em) {$\timesl$};
\node (iv) at (17em,8em) {$np\backslash s$};
\node (subj) at (11em,7.85em) {$np$};
\node[tns] (t2) at (14em,5em) {};
\node (t2lab) at (14em,5em) {+};
\node (s1) at (14em,2em) {$s$};
\draw (t2) -- (s1);
\draw (t2) -- (subj);
\draw (t2) -- (iv);
\draw (t1) -- (iv);
\draw (t1) -- (wtv);
\draw (t1) -- (obj);
\node (gq) at (17em,-4em) {$(s\upl np)\downl s$};
\node (ql) at (17em,-2.5em) {$\textit{everyone}$};
\draw (17em,-3.1em) -- (17em,-3.5em);
\node (snp) at (11em,-4em) {$s\upl np$};
\node (sg) at (14em,-10em) {$s$};
\node[tns] (t3) at (14em,-7em) {};
\node (t3lab) at (14em,-7em) {$\timesl$};
\draw (t3) -- (gq);
\draw (t3) -- (snp);
\draw (t3) -- (sg);
\node[par] (par) at (14em,-1em) {};
\node (parlab) at (14em,-1em) {\textcolor{white}{$\upl$}};
\draw[>=latex,->] (par) -- (snp);
\draw (par) -- (s1);
\draw (par) [in=30,out=-30] to (obj);
\node (mlab) at (33em,7.5em) {$\textit{Mary}$};
\draw (33em,6.6em) -- (33em,7.0em);
\node (wtv) at (36em,12em) {$\centerdot$};
\draw (36em,12.6em) -- (36em,13.0em);
\node (rang) at (34.5em,15em) {$\textit{rang}$};
\node (one) at (36.3em,15.2em) {$\one$};
\node (up) at (37.5em,15.05em) {$\textit{up}$};
\draw (34.5em,14em) -- (34.5em,14.5em);
\draw (36.3em,14em) -- (36.3em,14.5em);
\draw (37.5em,14em) -- (37.5em,14.5em);
\draw (34.5em,14em) -- (37.5em,14em) -- (37.5em,13em) -- (34.5em,13em) -- cycle;
\node (obj) at (42em,12em) {$\centerdot$};
\node[tns] (t1) at (39em,9em) {};
\node (t1lab) at (39em,9em) {$\timesl$};
\node (iv) at (39em,6em) {$\centerdot$};
\node (subj) at (33em,6em) {$\centerdot$};
\draw (33em,3.5em) -- (33em,4.5em) -- (39em,4.5em) -- (39em,3.5em) -- cycle; 
\node (s1) at (36em,2em) {$\centerdot$};
\draw (36em,3.5em) -- (s1);
\draw (33em,4.5em) -- (subj);
\draw (39em,4.5em) -- (iv);
\draw (t1) -- (iv);
\draw (t1) -- (wtv);
\draw (t1) -- (obj);
\node (gq) at (39em,-4em) {$\centerdot$};
\node (ql) at (39em,-2.5em) {$\textit{everyone}$};
\draw (39em,-3.1em) -- (39em,-3.5em);
\node (snp) at (33em,-4em) {$\centerdot$};
\node (sg) at (36em,-10em) {$s$};
\node[tns] (t3) at (36em,-7em) {};
\node (t3lab) at (36em,-7em) {$\timesl$};
\draw (t3) -- (gq);
\draw (t3) -- (snp);
\draw (t3) -- (sg);
\node[par] (par) at (36em,-1em) {};
\node (parlab) at (36em,-1em) {\textcolor{white}{$\upl$}};
\draw[>=latex,->] (par) -- (snp);
\draw (par) -- (s1);
\draw (par) [in=30,out=-30] to (obj);
\node at (28em,4em) {$\apsarrow$};
\end{tikzpicture}
\caption{Proof structure (left) and abstract proof structure (right) for the unfolding of ``Mary rang everyone up'' shown in Figure~\ref{fig:unfold}.}
\label{fig:pstoaps}
\end{figure}
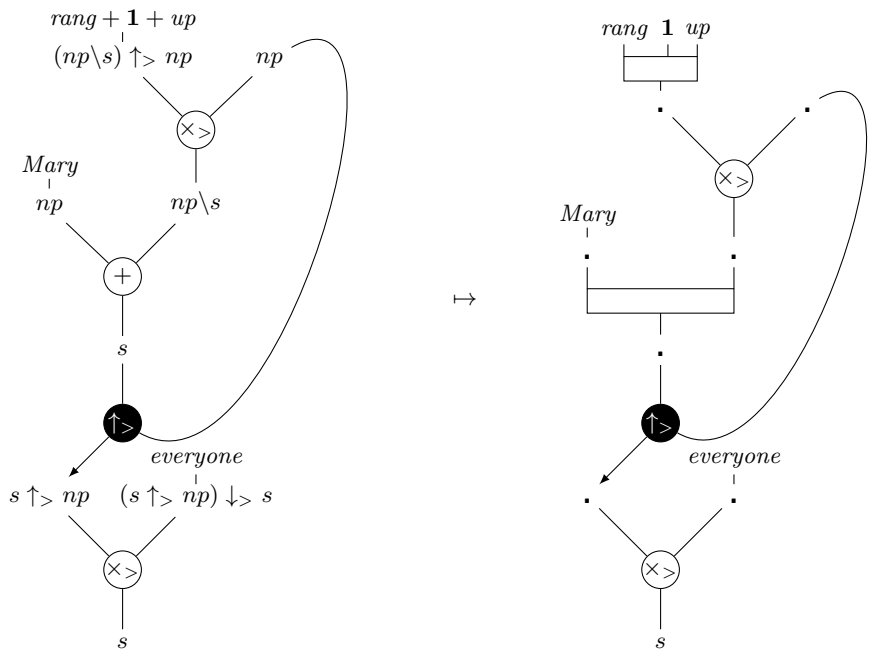

Figure~\ref{fig:unfold} gives an unfolding for the sentence ``Mary rang everyone up'', a sentence with the discontinuous idiom ``rang up'' and a
non-peripheral quantifier ``everyone'', following lexical entries of \citeasnoun{mvf11displacement}. Figure~\ref{fig:pstoaps} shows (on the left of the figure) one of the possibilities for connecting the atomic formulas.

Not all proof structures correspond to natural deduction proofs. Proof
structures which correspond to natural deduction proofs are
\emph{proof nets}. Of course, defining proof nets this way is not very
satisfactory: we want to have a condition which, given a proof
structure tells us whether or not this proof structure is a proof
net using only properties of the proof structure itself.

\subsection{Abstract Proof Structures}

The general strategy we follow to define a correctness criterion for proof structures is as follows: we first simplify by removing some of the information which is irrelevant for deciding correctness to obtain abstract proof structures, then specify the correctness condition
on these abstract proof structures, using a graph contraction criterion, generalizing the proof nets of the Lambek calculus from \citeasnoun{mp}.


\paragraph{Tensor trees and combs} A \emph{tensor tree} is a connected, acyclic set of tensor links (to be more precise, the underlying undirected graph must by acyclic and connected). A single vertex is a tensor tree. Given an (abstract) proof structure,
its tensor trees are the maximal substructures which are tensor trees; this is simply the forest we obtain when we remove all par links from a proof structure. The proof structure of Figure~\ref{fig:pstoaps} has two tensor trees.

A \emph{comb} is a link with any number of premisses and a single conclusion. None of the premisses of the comb can be identical to its conclusion. The general conditions
on links prevent premisses from being connected more than once as a premiss of a comb. The premisses of combs, as links in general, are linearly ordered. Premisses of a comb can be hypotheses of the proof structure, the conclusions of a link or
the special constant $\one$. 
The \emph{sort} of a comb, that is the sort assigned to its conclusion, is the sum of the sorts of its premisses (the constant $\one$ is of sort 1). Combs play the same role as tensor trees do for \citeasnoun{mp}: they
allow us to go back a forth between sequents $\Gamma \vdash C$ and combs with premisses $\Gamma$ and conclusion $C$. Given a comb, we will refer to subsequences of its
premisses as prefixes,
postfixes, etc., and assign them sorts as well.

\editout{
\begin{tikzpicture}
\node (p) at (2em,2em) {$A$};
\draw (p) -- (2em,0.5em);
\node at (10em,0.25em) {$\centerdot$};
\draw (8em,1.5em) -- (8em,2.5em) -- (12em,2.5em) -- (12em,1.5em) -- cycle;
\draw (9em,2.5em) -- (9em,3em);
\draw (11em,2.5em) -- (11em,3em);
\node (a1) at (8em,3.3em) {$p_0$};
\node (an) at (12em,3.3em) {$p_n$};
\node at (9em,3.5em) {$\one$};
\node (dots) at (10em,3.5em) {$\ldots$};
\draw (10em,2.5em) -- (10em,3em);
\node at (11em,3.5em) {$\one$};
\draw (10em,1.5em) -- (10em,0.5em);
\draw (10em,-0.5em) -- (10em,-1.5em);
\draw (8em,2.5em) -- (8em,3.0em);
\draw (12em,2.5em)  -- (12em,3.0em);
\end{tikzpicture}
}

\paragraph{Translating a proof structure to an abstract proof structure}
To translate a proof structure $\mathcal{P}$ to an abstract proof structure $\mathcal{A}$, we define a function, $\mathcal{P} \apsarrow \mathcal{A}$, which replaces ``+''
links by 2-premiss combs as follows

\begin{figure}
\begin{tikzpicture}
\node (p) at (-14em,2em) {$A$};
\draw (p) -- (-14em,0.5em);
\node at (-6em,0.25em) {$\centerdot$};
\draw (-8em,1.5em) -- (-8em,2.5em) -- (-4em,2.5em) -- (-4em,1.5em) -- cycle;
\draw (-7em,2.5em) -- (-7em,3em);
\draw (-5em,2.5em) -- (-5em,3em);
\node (a1) at (-8em,3.3em) {$p_0$};
\node (an) at (-4em,3.3em) {$p_n$};
\node at (-7em,3.5em) {$\one$};
\node (dots) at (-6em,3.5em) {$\ldots$};
\draw (-6em,2.5em) -- (-6em,3em);
\node at (-5em,3.5em) {$\one$};
\draw (-6em,1.5em) -- (-6em,0.5em);
\draw (-6em,-0.5em) -- (-6em,-1.5em);
\draw (-8em,2.5em) -- (-8em,3.0em);
\draw (-4em,2.5em)  -- (-4em,3.0em);
\node (t1) at (-9em,7em) {Lexical inputs};
\node (p) at (2em,2em) {$A$};
\draw (p) -- (2em,3.5em);
\draw (p) -- (2em,0.5em);
\node at (10em,0.25em) {$\centerdot$};
\draw (8em,1.5em) -- (8em,2.5em) -- (12em,2.5em) -- (12em,1.5em) -- cycle;
\draw (8em,4.0em) [in=270,out=90] to (10em,6.5em);
\draw (12em,4.0em) [in=270,out=90] to (10em,6.5em);
\draw (8em,2.5em) -- (8em,3em);
\draw (9em,2.5em) -- (9em,3em);
\draw (11em,2.5em) -- (11em,3em);
\draw (12em,2.5em) -- (12em,3em);
\node  at (8em,3.5em) {$\centerdot$};
\node  at (12em,3.5em) {$\centerdot$};
\node at (9em,3.5em) {$\one$};
\node (dots) at (10em,3.5em) {$\ldots$};
\node at (11em,3.5em) {$\one$};
\draw (10em,1.5em) -- (10em,0.5em);
\draw (dots) -- (10em,6.5em);
\draw (10em,-0.5em) -- (10em,-1.5em);
\draw (10em,2.5em) -- (10em,3.0em);
\node (t1) at (6em,7em) {Auxiliary inputs};
\node at (6em,2em) {$\apsarrow$};
\node at (-10em,2em) {$\apsarrow$};
\end{tikzpicture}
\begin{tikzpicture}
\node (p) at (-14em,2em) {$A$};
\draw (p) -- (-14em,3.5em);
\node (t1) at (-9em,7em) {Conclusion};
\node (p) at (-6em,2em) {$A$};
\draw (p) -- (-6em,3.5em);
\node (p) at (2em,2em) {$A$};
\draw (p) -- (2em,3.5em);
\draw (p) -- (2em,0.5em);
\node (p) at (10em,2.0em) {$\centerdot$};
\draw (p) -- (10em,3.5em);
\draw (p) -- (10em,0.5em);
\node (t1) at (6em,7em) {Other internal nodes};
\node at (6em,2em) {$\apsarrow$};
\node at (-10em,2em) {$\apsarrow$};
\end{tikzpicture}
\caption{Conversion to abstract proof structures for vertices/formulas.}
\label{fig:vaps}
\end{figure}
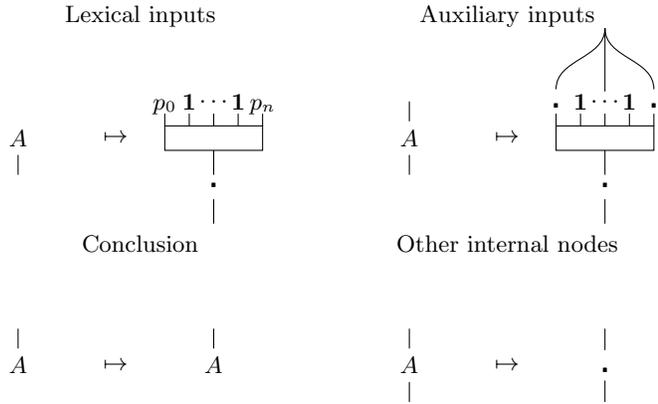

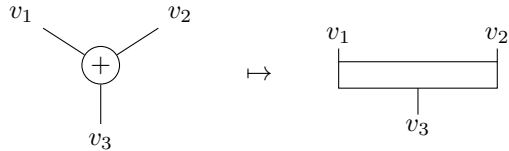
\begin{tikzpicture}
\node (ab) at (3em,0.0em) {$v_3$};
\node (a) at (0,4.8em) {$v_1$};
\node (b) at (6em,4.8em) {$v_2$};
\node[tns] (c) at (3em,2.868em) {};
\node (clab) at (3em,2.868em) {+};
\draw (c) -- (ab);
\draw (c) -- (a);
\draw (c) -- (b);
\draw (12em,2em) -- (18em,2em) -- (18em,3em) -- (12em,3em) -- cycle;
\draw (15em,2em) -- (15em,1em);
\draw (12em,3em) -- (12em,3.5em);
\draw (18em,3em) -- (18em,3.5em);
\node at (12em,4.0em) {$v_1$};
\node at (18em,4.0em) {$v_2$};
\node at (15em,0.5em) {$v_3$};
\node at (9em,2.5em) {$\apsarrow$};
\end{tikzpicture} 

\noindent which leaves all other links the same 
 and which replaces the
vertices/formulas of $\mathcal{P}$ as shown in Figure~\ref{fig:vaps}. The only slight complication is for the input formulas (lexical our auxiliary).
Proof structures are defined as ways of connecting formulas, but for formulating correctness we need to know about the strings
denoted by these formulas, for example, about their position relative to other formulas, separator symbols or the left/rightmost position. Another way of seeing this is that
we need to replace sorted variables $\alpha$ (such as those assigned to hypotheses) by variables $p_0 + \one + \ldots + \one + p_n$, with each $p_i$ of sort 0 (such
a strategy is already implicitly used for the natural deduction rules for $/I$, $\backslash I$, $\bullet E$, $\upk I$, $\downk I$ and $\prodk E$, that is the natural deduction
rules corresponding to the par links).  As shown in Figure~\ref{fig:vaps},
auxiliary inputs separate the path leaving the par link by adding  $n$ new subpaths (this appears somewhat odd, but is required for the correct behaviour of the contractions when sorts
are greater than 0, as we will see below). Because of the sorts of the formulas, the par links for $\downk$ and $\prodk$ necessarily involve at least one such split, 
though the other par links need not.

\subsection{Contractions}

\paragraph{Structural contractions}


\begin{figure}
\begin{tikzpicture}
\draw (5em,2em) -- (5em,3em) -- (15em,3em) -- (15em,2em)  -- cycle;
\node (di) at (10em,4.3em) {$\centerdot$};
\node at (6.25em,4em) {$\alpha_1$};
\node at (6.25em,3.3em) {$\scriptscriptstyle{\ldots}$};
\node at (13.5em,4em) {$\alpha_2$};
\node at (13.5em,3.3em) {$\scriptscriptstyle{\ldots}$};
\draw (5em,3em) -- (5em,3.5em);
\draw (8em,3em) -- (8em,3.5em);
\draw (10em,3em) -- (10em,3.5em);
\draw (12em,3em) -- (12em,3.5em);
\draw (15em,3em) -- (15em,3.5em);
\draw (10em,2em) -- (10em,1em);
\draw (10em,5em) -- (10em,6em);
\draw (8em,6em) -- (12em,6em) -- (12em,7em) -- (8em,7em) -- cycle;
\node at (10em,7.3em) {$\scriptscriptstyle{\ldots}$};
\node at (10em,8em) {$\beta$};
\draw (8em,7em) -- (8em,7.5em);
\draw (12em,7em) -- (12em,7.5em);
\editout{
\draw (21em,2em) -- (34em,2em) -- (34em,3em) -- (21em,3em) -- cycle;
\node at (21.25em,4em) {$\alpha_1$};
\node at (21.25em,3.3em) {$\scriptscriptstyle{\ldots}$};
\node at (27em,4em) {$\beta$};
\node at (27em,3.3em) {$\scriptscriptstyle{\ldots}$};
\node at (32.5em,4em) {$\alpha_2$};
\node at (32.5em,3.3em) {$\scriptscriptstyle{\ldots}$};
\draw (21em,3em) -- (21em,3.5em);
\draw (23em,3em) -- (23em,3.5em);
\draw (25em,3em) -- (25em,3.5em);
\draw (29em,3em) -- (29em,3.5em);
\draw (31em,3em) -- (31em,3.5em);
\draw (34em,3em) -- (34em,3.5em);
\draw (27em,2em) -- (27em,1em);
}%
\node (arrow2) at (17.6em,2.8em) {$\overset{[+]}{\rightarrow}$};
\draw (21em,2em) -- (21em,3em) -- (29em,3em) -- (29em,2em)  -- cycle;
\draw (21em,3em) -- (21em,3.5em);
\draw (23em,3em) -- (23em,3.5em);
\draw (24em,3em) -- (24em,3.5em);
\draw (26em,3em) -- (26em,3.5em);
\draw (27em,3em) -- (27em,3.5em);
\draw (29em,3em) -- (29em,3.5em);
\node at (22em,3.3em) {$\scriptscriptstyle{\ldots}$};
\node (a1) at (22em,4.0em) {$\alpha_1$};
\node at (25em,3.3em) {$\scriptscriptstyle{\ldots}$};
\node (b) at (25em,4.0em) {$\beta$};
\node at (28em,3.3em) {$\scriptscriptstyle{\ldots}$};
\node (a2) at (28em,4.0em) {$\alpha_2$};
\draw (25em,2em) -- (25em,1em);
\node (v1) at (10em,0em) {$v$};
\node (v2) at (25em,0em) {$v$};
\end{tikzpicture}
\bigskip
\begin{tikzpicture}
\draw (0em,0em) -- (0em,1em) -- (9em,1em) -- (9em,0em)  -- cycle;
\draw (0em,1em) -- (0em,1.5em);
\draw (3em,1em) -- (3em,1.5em);
\draw (6em,1em) -- (6em,1.5em);
\draw (4.5em,0em) -- (4.5em,1.5em);
\draw (9em,1em) -- (9em,1.5em);
\node (a1) at (1.5em,2.0em) {$\alpha_1$};
\node at (1.5em,1.3em) {$\scriptscriptstyle{\ldots}$};
\node (one) at (4.5em,2.2em) {$\one$};
\node (a2) at (7.5em,2.0em) {$\alpha_2$};
\node at (7.5em,1.3em) {$\scriptscriptstyle{\ldots}$};
\draw (10em,0em) -- (10em,1em) -- (14em,1em) -- (14em,0em)  -- cycle;
\draw (10em,1em) -- (10em,1.5em);
\draw (14em,1em) -- (14em,1.5em);
\node (a1) at (12em,2.0em) {$\beta$};
\node at (12em,1.3em) {$\scriptscriptstyle{\ldots}$};
\node[tns] (t) at (8em,-3em) {};
\node (tlab) at (8em,-3em) {$\timesk$};
\draw (t) -- (4.5em,-1.3em);
\draw (t) -- (12em,-1.3em);
\draw (t) -- (8em,-5em);
\node at (4.5em,-0.8em) {$\centerdot$};
\node at (12em,-0.8em) {$\centerdot$};
\draw (4.5em,-0.5em) -- (4.5em,0em);
\draw (12em,-0.5em) -- (12em,0em);
\draw (18em,0em) -- (18em,1em) -- (26em,1em) -- (26em,0em)  -- cycle;
\draw (18em,1em) -- (18em,1.5em);
\draw (20em,1em) -- (20em,1.5em);
\draw (21em,1em) -- (21em,1.5em);
\draw (23em,1em) -- (23em,1.5em);
\draw (24em,1em) -- (24em,1.5em);
\draw (26em,1em) -- (26em,1.5em);
\node at (19em,1.3em) {$\scriptscriptstyle{\ldots}$};
\node (a1) at (19em,2.0em) {$\alpha_1$};
\node at (22em,1.3em) {$\scriptscriptstyle{\ldots}$};
\node (b) at (22em,2.0em) {$\beta$};
\node at (25em,1.3em) {$\scriptscriptstyle{\ldots}$};
\node (a2) at (25em,2.0em) {$\alpha_2$};
\draw (22em,0em) -- (22em,-1em);
\node (arrow2) at (16.3em,0.8em) {$\overset{[\timesk]}{\rightarrow}$};
\node (v1) at (8em,-6em) {$v$};
\node (v2) at (22em,-2em) {$v$};
\end{tikzpicture}
\vspace{-\baselineskip}
\caption{Structural contractions}
\label{fig:sr}
\end{figure}
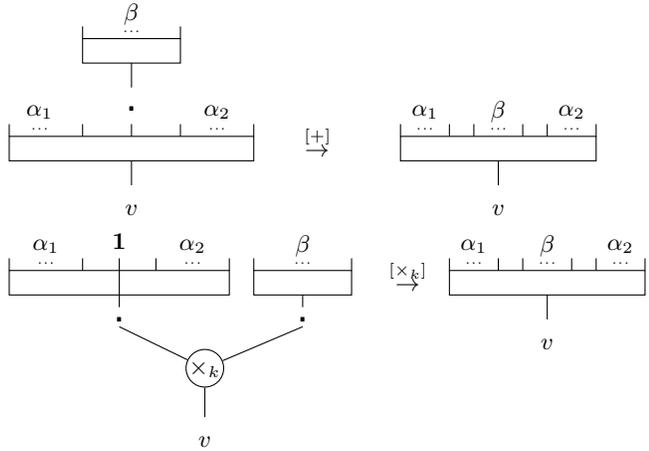

Figure~\ref{fig:sr} shows the structural contractions. The ``+'' contraction composes two combs, combining the premisses by a simple left-to-right traversal.
It is worth mentioning some immediate corollaries of this contraction here: first, we simply eliminate trivial combs (containing a single premiss and a single
conclusion, that is, when $\beta$ contains only a single premiss)  when their conclusion is the premiss of another comb, 
and, second, the structural contractions contract tensor
trees to unique combs (this is no longer guaranteed once we add the synthetic connectives, as discussed in Section~\ref{sec:extension}).

The wrap operation ``$\timesk$'' reflects the wrap operation on strings on the level of abstract proof structures, it inserts $\beta$ at the separator indicated by $k$:
if $k$ is ``$>$'', the $\alpha_1$ must be of sort 0 (we replace the first separator by $\beta$) and if $k$ is ``$<$'' $\alpha_2$ must be of sort 0 (we replace the last separator by $\beta$). 
 
Note that  $\alpha_1$, $\alpha_2$ and $\beta$ are allowed to have zero premisses.

\paragraph{Logical contractions} The logical contractions ensure the logical symmetry of the connectives in the calculus. Each par rule has its own contraction. The
contraction for $\backslash$, shown below, essentially checks whether the string term of the premiss is equivalent to $p_0 + \one + \ldots + \one + p_n + \beta$ and withdraws
the hypothesis $p_0 + \one + \ldots + \one + p_n$ (where $n$ is the sort
of the withdrawn formula in the corresponding $\backslash I$ rule) to reduce to $\beta$ (the $/$ contraction is left-right symmetric).

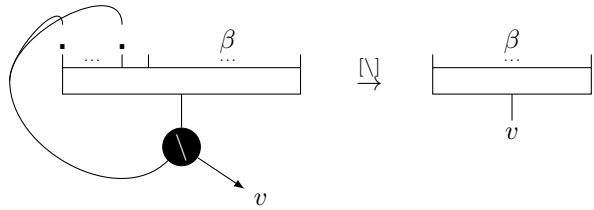
\begin{tikzpicture}
\draw (0em,0em) -- (0em,1em) -- (9em,1em) -- (9em,0em)  -- cycle;
\draw (0em,1em) -- (0em,1.5em);
\draw (2.25em,1em) -- (2.25em,1.5em);
\draw (3.25em,1em) -- (3.25em,1.5em);
\draw (9em,1em) -- (9em,1.5em);
\node (a1l) at (0em,2.0em) {$\centerdot$};
\node (a1r) at (2.25em,2.0em) {$\centerdot$};
\node at (1.125em,1.3em) {$\scriptscriptstyle{\ldots}$};
\node at (6.25em,1.3em) {$\scriptscriptstyle{\ldots}$};
\node (b) at (6.25em,2.0em) {$\beta$};
\node[par] (p) at (4.5em,-2em) {};
\node (plab) at (4.5em,-2em) {\textcolor{white}{$\backslash$}};
\draw (p) -- (4.5em,0em);
\node (r) at (7.5em,-4em) {$v$};
\draw[>=latex,<-] (r) -- (p);
\draw (p) to [in=270,out=225] (-2em,0.5em);
\draw (-2em,0.5em) to [in=90,out=90] (a1l);
\draw (-2em,0.5em) to [in=90,out=90] (a1r);
%
\draw (14em,0em) -- (14em,1em) -- (20em,1em) -- (20em,0em)  -- cycle;
\draw (17.0em,0em) -- (17.0em,-1em);
\node (a1) at (17.0em,2.0em) {$\beta$};
\node at (17em,1.3em) {$\scriptscriptstyle{\ldots}$};
\node (arrow2) at (11.6em,0.8em) {$\overset{[\backslash]}{\rightarrow}$};
\draw (14em,1em) -- (14em,1.5em);
\draw (20em,1em) -- (20em,1.5em);
\node (v2) at (17em,-1.5em) {$v$};
\end{tikzpicture}

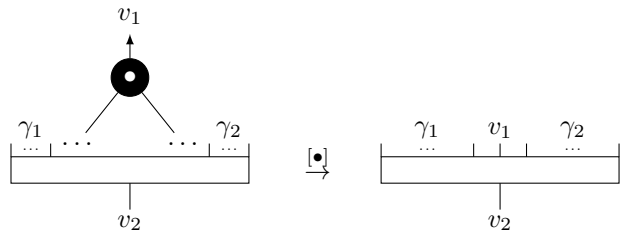
\begin{tikzpicture}
\draw (0em,0em) -- (0em,1em) -- (9em,1em) -- (9em,0em)  -- cycle;
\node (g1) at (0.75em,2.0em) {$\gamma_1$};
\node (g2) at (8.25em,2.0em) {$\gamma_2$};
\node (g1) at (0.75em,1.3em) {$\scriptscriptstyle{\ldots}$};
\node (g2) at (8.25em,1.3em) {$\scriptscriptstyle{\ldots}$};
\draw (0em,1em) -- (0em,1.5em);
\draw (1.5em,1em) -- (1.5em,1.5em);
\draw (9em,1em) -- (9em,1.5em);
\draw (7.5em,1em) -- (7.5em,1.5em);
\node (a1) at (2.5em,1.5em) {$\ldots$};
\node (a2) at (6.5em,1.5em) {$\ldots$};
\draw (4.5em,0em) -- (4.5em,-1em);
\node[par] (p) at (4.5em,4.0em) {};
\node (plab) at (4.5em,4.0em) {\textcolor{white}{$\bullet$}};
\node (top) at (4.5em,6.3em) {$v_1$};
\draw[>=latex,<-] (top) -- (p);
\draw (p) -- (a2);
\draw (p) -- (a1);
\draw (14em,0em) -- (14em,1em) -- (23em,1em) -- (23em,0em)  -- cycle;
\draw (14em,1em) -- (14em,1.5em);
\draw (17.5em,1em) -- (17.5em,1.5em);
\draw (18.5em,1em) -- (18.5em,1.5em);
\draw (19.5em,1em) -- (19.5em,1.5em);
\draw (23em,1em) -- (23em,1.5em);
\draw (18.5em,0em) -- (18.5em,-1em);
\node (a1) at (15.75em,2.0em) {$\gamma_1$};
\node (a2) at (21.25em,2.0em) {$\gamma_2$};
\node (g1) at (15.75em,1.3em) {$\scriptscriptstyle{\ldots}$};
\node (g2) at (21.25em,1.3em) {$\scriptscriptstyle{\ldots}$};
\node (arrow2) at (11.6em,0.8em) {$\overset{[\bullet]}{\rightarrow}$};
\node (v1) at (4.5em,-1.5em) {$v_2$};
\node (v2) at (18.5em,-1.5em) {$v_2$};
\node (w2) at (18.5em,2.0em) {$v_1$};
\end{tikzpicture}

The
contraction for $\upk$ essentially checks that its auxiliary input is an infix of the appropriate sort.

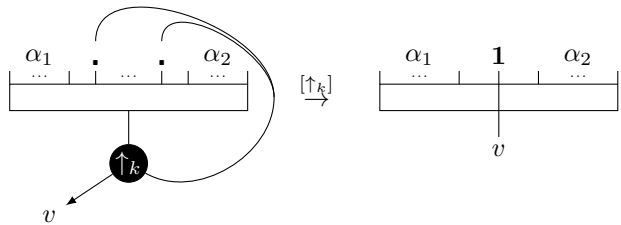
\begin{tikzpicture}
\draw (0em,0em) -- (0em,1em) -- (9em,1em) -- (9em,0em)  -- cycle;
\draw (0em,1em) -- (0em,1.5em);
\draw (2.25em,1em) -- (2.25em,1.5em);
\draw (3.25em,1em) -- (3.25em,1.5em);
\draw (5.75em,1em) -- (5.75em,1.5em);
\draw (6.75em,1em) -- (6.75em,1.5em);
\draw (9em,1em) -- (9em,1.5em);
\node (a1) at (1.125em,2.0em) {$\alpha_1$};
\node at (1.125em,1.3em) {$\scriptscriptstyle{\ldots}$};
\node (b1) at (3.25em,2.0em) {$\centerdot$};
\node (b2) at (5.75em,2.0em) {$\centerdot$};
\node at (4.5em,1.3em) {$\scriptscriptstyle{\ldots}$};
\node (a2) at (7.875em,2.0em) {$\alpha_2$};
\node at (7.875em,1.3em) {$\scriptscriptstyle{\ldots}$};
\node[par] (p) at (4.5em,-2em) {};
\node (plab) at (4.5em,-2em) {\textcolor{white}{$\upk$}};
\draw (p) -- (4.5em,0em);
\node (r) at (1.5em,-4em) {$v$};
\draw[>=latex,<-] (r) -- (p);
\draw (p) to [in=270,out=-30] (10em,0.5em);
\draw (10em,0.5em) to [in=90,out=90] (b1);
\draw (10em,0.5em) to [in=90,out=90] (b2);
\draw (14em,0em) -- (14em,1em) -- (23em,1em) -- (23em,0em)  -- cycle;
\draw (18.5em,-1em) -- (18.5em,1.5em);
\draw (14em,1em) -- (14em,1.5em);
\draw (17em,1em) -- (17em,1.5em);
\draw (20em,1em) -- (20em,1.5em);
\draw (23em,1em) -- (23em,1.5em);
\node (a1) at (15.5em,2.0em) {$\alpha_1$};
\node at (15.5em,1.3em) {$\scriptscriptstyle{\ldots}$};
\node (one) at (18.5em,2.1em) {$\one$};
\node (a2) at (21.5em,2.0em) {$\alpha_2$};
\node at (21.5em,1.3em) {$\scriptscriptstyle{\ldots}$};
\node (arrow2) at (11.6em,0.8em) {$\overset{[\upk]}{\rightarrow}$};
\node (v) at (18.5em,-1.5em) {$v$};
\end{tikzpicture}

Depending on $k$, there are restrictions of the sorts: for ``$>$'', $\alpha_1$ must be of sort 0 (that is, all premisses of the comb to the left $\beta$ are of sort 0),
for ``$<$'', $\alpha_2$ must be of sort 0
(that is, all premisses of the comb to the right $\beta$ are of sort 0), for $k = n$,
$\alpha_1 $ is a prefix of sort $n-1$ (that is, the sorts of the premisses of the comb to the left $\beta$ sum to $n-1$).

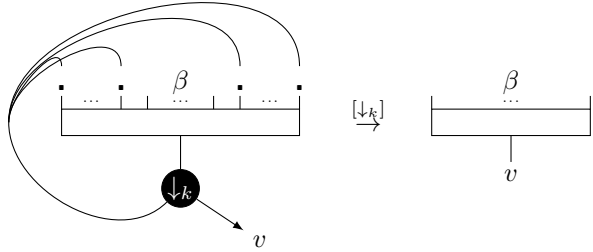
\begin{tikzpicture}
\draw (0em,0em) -- (0em,1em) -- (9em,1em) -- (9em,0em)  -- cycle;
\draw (0em,1em) -- (0em,1.5em);
\draw (2.25em,1em) -- (2.25em,1.5em);
\draw (3.25em,1em) -- (3.25em,1.5em);
\draw (5.75em,1em) -- (5.75em,1.5em);
\draw (6.75em,1em) -- (6.75em,1.5em);
\draw (9em,1em) -- (9em,1.5em);
\node (a1l) at (0em,2.0em) {$\centerdot$};
\node (a1r) at (2.25em,2.0em) {$\centerdot$};
\node (a2l) at (6.75em,2.0em) {$\centerdot$};
\node (a2r) at (9.0em,2.0em) {$\centerdot$};
\node at (1.125em,1.3em) {$\scriptscriptstyle{\ldots}$};
\node at (7.8725em,1.3em) {$\scriptscriptstyle{\ldots}$};
\node at (4.5em,1.3em) {$\scriptscriptstyle{\ldots}$};
\node (b) at (4.5em,2.0em) {$\beta$};
\node[par] (p) at (4.5em,-2em) {};
\node (plab) at (4.5em,-2em) {\textcolor{white}{$\downk$}};
\draw (p) -- (4.5em,0em);
\node (r) at (7.5em,-4em) {$v$};
\draw[>=latex,<-] (r) -- (p);
\draw (p) to [in=270,out=225] (-2em,0.5em);
\draw (-2em,0.5em) to [in=90,out=90] (a1l);
\draw (-2em,0.5em) to [in=90,out=90] (a1r);
\draw (-2em,0.5em) to [in=90,out=90] (a2l);
\draw (-2em,0.5em) to [in=90,out=90] (a2r);
\draw (14em,0em) -- (14em,1em) -- (20em,1em) -- (20em,0em)  -- cycle;
\draw (17.0em,0em) -- (17.0em,-1em);
\node (a1) at (17.0em,2.0em) {$\beta$};
\node at (17em,1.3em) {$\scriptscriptstyle{\ldots}$};
\node (arrow2) at (11.6em,0.8em) {$\overset{[\downk]}{\rightarrow}$};
\draw (14em,1em) -- (14em,1.5em);
\draw (20em,1em) -- (20em,1.5em);
\node (v) at (17em,-1.5em) {$v$};
\end{tikzpicture}

If $k$ is ``$>$'', the premisses to the left of $\beta$ are of sort 0. If $k$ is  ``$<$'', the premisses to the right of $\beta$ are of sort 0. If $k = n$, the premisses to the left of
$\beta$ are of sort $n-1$. 
This contraction looks odd until we realize that we are dealing with a circumfix operation and that, as a consequence the subformula $A$ of a formula $A\downk B$ denotes a
discontinuous circumfix with corresponding string $\alpha_1 + \one + \alpha_2$ (look back to the introduction rule for $\downk$ on the top right of Figure~\ref{fig:ndlw} for comparison).

The contraction for $\prodk$ generalizes the contraction for $\bullet$. Whereas the contraction for $A\bullet B$ verifies the strings of the subformulas $A$ and $B$ are
adjacent, the contraction for $A\prodk B$ verifies whether the string $\alpha_1\dplus\one\dplus\alpha_2$ of $A$ is circumfixed around the string $\beta$ of $B$. The sorts
of $\alpha_1$, $\alpha_2$ and $\beta$ depend on $k$ and on the sorts of $A$ and $B$, exactly as for the other rules (in the rule below, the labels $\alpha_1$, $\alpha_2$ and $\beta$
represent sequences of vertices which are premisses of the comb and conclusions of the $\prodk$ rule).

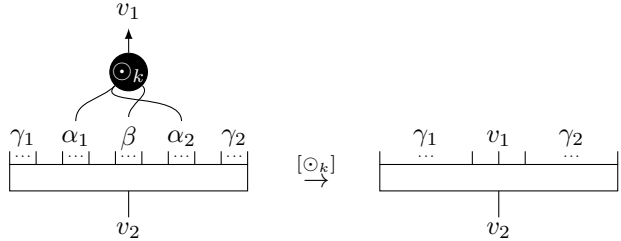
\begin{tikzpicture}
\draw (0em,0em) -- (0em,1em) -- (9em,1em) -- (9em,0em)  -- cycle;
\draw (0em,1em) -- (0em,1.5em);
\draw (1em,1em) -- (1em,1.5em);
\draw (2em,1em) -- (2em,1.5em);
\draw (3em,1em) -- (3em,1.5em);
\draw (4em,1em) -- (4em,1.5em);
\draw (5em,1em) -- (5em,1.5em);
\draw (6em,1em) -- (6em,1.5em);
\draw (7em,1em) -- (7em,1.5em);
\draw (8em,1em) -- (8em,1.5em);
\draw (9em,1em) -- (9em,1.5em);
\node at (0.5em,1.3em) {$\scriptscriptstyle{\ldots}$};
\node at (2.5em,1.3em) {$\scriptscriptstyle{\ldots}$};
\node at (4.5em,1.3em) {$\scriptscriptstyle{\ldots}$};
\node at (6.5em,1.3em) {$\scriptscriptstyle{\ldots}$};
\node at (8.5em,1.3em) {$\scriptscriptstyle{\ldots}$};
\node (g1) at (0.5em,2.0em) {$\gamma_1$};
\node (g2) at (8.5em,2.0em) {$\gamma_2$};
\node (a1) at (2.5em,2.0em) {$\alpha_1$};
\node (b) at (4.5em,2.0em) {$\beta$};
\node (a2) at (6.5em,2.0em) {$\alpha_2$};
\draw (4.5em,0em) -- (4.5em,-1em);
\node[par] (p) at (4.5em,4.5em) {};
\node (plab) at (4.5em,4.5em) {\textcolor{white}{$\prodk$}};
\node (top) at (4.5em,6.8em) {$v_1$};
\draw[>=latex,<-] (top) -- (p);
\draw (p) to [in=90,out=315] (b);
\draw (p) to [in=90,out=225] (a1);
\draw (p) to [in=90,out=225] (a2);
\draw (14em,0em) -- (14em,1em) -- (23em,1em) -- (23em,0em)  -- cycle;
\draw (14em,1em) -- (14em,1.5em);
\draw (17.5em,1em) -- (17.5em,1.5em);
\draw (18.5em,1em) -- (18.5em,1.5em);
\draw (19.5em,1em) -- (19.5em,1.5em);
\draw (23em,1em) -- (23em,1.5em);
\draw (18.5em,0em) -- (18.5em,-1em);
\node (a1) at (15.75em,2.0em) {$\gamma_1$};
\node (a2) at (21.25em,2.0em) {$\gamma_2$};
\node at (15.75em,1.3em) {$\scriptscriptstyle{\ldots}$};
\node at (21.25em,1.3em) {$\scriptscriptstyle{\ldots}$};
\node (arrow2) at (11.6em,0.8em) {$\overset{[\prodk]}{\rightarrow}$};
\node (v2) at (4.5em,-1.5em) {$v_2$};
\node (v1r) at (18.5em,2.0em) {$v_1$};
\node (v2r) at (18.5em,-1.5em) {$v_2$};
\end{tikzpicture}

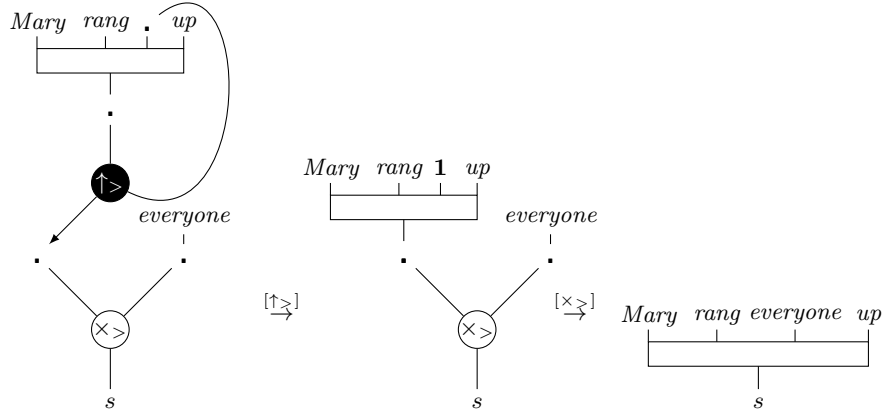
\begin{figure}[tb]
\begin{tikzpicture}
\node (mlab) at (33em,5.55em) {$\textit{Mary}$};
\node (rang) at (35.8em,5.5em) {$\textit{rang}$};
\node (obj) at (37.5em,5.5em) {$\centerdot$};
\node (up) at (39.0em,5.50em) {$\textit{up}$};
\draw (33em,3.5em) -- (33em,4.5em) -- (39em,4.5em) -- (39em,3.5em) -- cycle; 
\node (s1) at (36em,2em) {$\centerdot$};
\draw (36em,3.5em) -- (s1);
\draw (33em,4.5em) -- (33em,5.0em);
\draw (35.8em,4.5em) -- (35.8em,5.0em);
\draw (37.5em,4.5em) -- (37.5em,5.0em);
\draw (39em,4.5em) -- (39em,5.0em);
\node (gq) at (39em,-4em) {$\centerdot$};
\node (ql) at (39em,-2.5em) {$\textit{everyone}$};
\draw (39em,-3.1em) -- (39em,-3.5em);
\node (snp) at (33em,-4em) {$\centerdot$};
\node (sg) at (36em,-10em) {$s$};
\node[tns] (t3) at (36em,-7em) {};
\node (t3lab) at (36em,-7em) {$\timesl$};
\draw (t3) -- (gq);
\draw (t3) -- (snp);
\draw (t3) -- (sg);
\node[par] (par) at (36em,-1em) {};
\node (parlab) at (36em,-1em) {\textcolor{white}{$\upl$}};
\draw[>=latex,->] (par) -- (snp);
\draw (par) -- (s1);
%
\draw (par) .. controls (42em,-4em) and (42em,9em) .. (obj);
\node (mlab) at (45em,-0.45em) {$\textit{Mary}$};
\node (rang) at (47.8em,-0.5em) {$\textit{rang}$};
\node (obj) at (49.5em,-0.35em) {$\one$};
\node (up) at (51.0em,-0.50em) {$\textit{up}$};
\draw (45em,-2.5em) -- (45em,-1.5em) -- (51em,-1.5em) -- (51em,-2.5em) -- cycle; 
\node (s1) at (48em,-4em) {$\centerdot$};
\draw (48em,-2.5em) -- (s1);
\draw (45em,-1.5em) -- (45em,-1.0em);
\draw (47.8em,-1.5em) -- (47.8em,-1.0em);
\draw (49.5em,-1.5em) -- (49.5em,-1.0em);
\draw (51em,-1.5em) -- (51em,-1.0em);
\node (gq) at (54em,-4em) {$\centerdot$};
\node (ql) at (54em,-2.5em) {$\textit{everyone}$};
\draw (54em,-3.1em) -- (54em,-3.5em);
\node (snp) at (48em,-4em) {$\centerdot$};
\node (sg) at (51em,-10em) {$s$};
\node[tns] (t3) at (51em,-7em) {};
\node (t3lab) at (51em,-7em) {$\timesl$};
\draw (t3) -- (gq);
\draw (t3) -- (snp);
\draw (t3) -- (sg);
\node (mlab) at (58em,-6.45em) {$\textit{Mary}$};
\node (rang) at (60.8em,-6.5em) {$\textit{rang}$};
\node (obj) at (64.0em,-6.45em) {$\textit{everyone}$};
\node (up) at (67.0em,-6.50em) {$\textit{up}$};
\draw (58em,-8.5em) -- (58em,-7.5em) -- (67em,-7.5em) -- (67em,-8.5em) -- cycle; 
\node (sg) at (62.5em,-10em) {$s$};
\draw (62.5em,-8.5em) -- (sg);
\draw (58em,-7.5em) -- (58em,-7.0em);
\draw (60.8em,-7.5em) -- (60.8em,-7.0em);
\draw (64em,-7.5em) -- (64em,-7.0em);
\draw (67em,-7.5em) -- (67em,-7.0em);
\node (arrow1) at (43em,-6em) {$\overset{[\upl]}{\rightarrow}$};
\node (arrow2) at (55em,-6em) {$\overset{[\timesl]}{\rightarrow}$};
\end{tikzpicture}
\caption{Contractions for the abstract proof structure for the sentence ``Mary rang everyone up'' shown on the right of Figure~\ref{fig:pstoaps}.}
\label{fig:contract}
\end{figure}

As an example of how we can use the contraction criterion to verify whether a proof structure is a proof net, the abstract proof structure of Figure~\ref{fig:pstoaps} can
be contracted first by using a ``+'' and a ``$\timesl$'' contraction to produce the abstract proof structure shown on the left of Figure~\ref{fig:contract}, then by
performing the ``$\upl$'' and ``$\timesl$'' contractions as indicated, giving a proof of ``Mary rang everyone up''.


\paragraph{Brief remarks on complexity}

It is easy to see the given contraction calculus is confluent and that
each of the contraction steps reduces the total number of links in the
structure. Therefore, even a naive implementation of this contraction
calculus checks whether or not a given abstract proof structure with
$n$ links contracts to a comb in $O(n^3)$ steps, simply by repeatedly
traversing the links in the graph to find contractible configurations
and contracting them once they are found (we can likely improve upon this worst case, but I will leave this to further research).

In particular, this shows NP-completeness of the Displacement
calculus. NP-hardness follows from NP-completeness of the Lambek calculus \cite{pentus06np} and we can show it is in NP since we can verify in polynomial time whether or not a
candidate proof (that is, a proof structure) in the Displacement calculus is a proof (that is, a proof net).


\subsection{Correctness of the calculus}

We show that the two definitions of proof net, contractibility and corresponding to a natural deduction proof coincide, thereby establishing that the contraction
criterion is correct.

\begin{lemma}\label{lem:sound} Let $\delta$ be a Displacement calculus natural deduction proof of $\alpha_1:A_1,\ldots,a_n:A_n \vdash \gamma:C$. There is a proof net with
  the same hypotheses whose abstract proof structure contracts to a $\gamma:C$. 
\end{lemma}

\paragraph{Proof} This is a simple induction on the length of the proof. Axioms $\alpha:A\vdash \alpha:A$ correspond directly to proof nets with the required
combs. Otherwise, we proceed by case analysis on the last rule of the proof. Each logical rule correspond to adding a link to the proof net(s) given by induction hypothesis and a contraction to the sequence of contractions for the abstract
proof structure. We show only the case for $\downl$. In this case, the last rule looks as follows.

\[
\infer[\downl I_i]{\gamma:A\downl C}{\infer*[\delta]{a\dplus\gamma\dplus\alpha:C}{[a\dplus\one\dplus\alpha:A]^i}}
\]

Induction hypothesis gives use a proof net of
$\Gamma,a\dplus\one\dplus\alpha:A \vdash
a\dplus\gamma\dplus\alpha:C$. That is we are in the situation shown
below, with the proof structure shown below on the right, the
corresponding abstract proof structure in the middle, for which we are
given a sequence of reductions $\rho$ to a comb
$a\dplus\gamma\dplus\alpha:C$. We have simply spelled out the
definition of proof net of $\Gamma,a\dplus\one\dplus\alpha:A \vdash
a\dplus\gamma\dplus\alpha:C$.

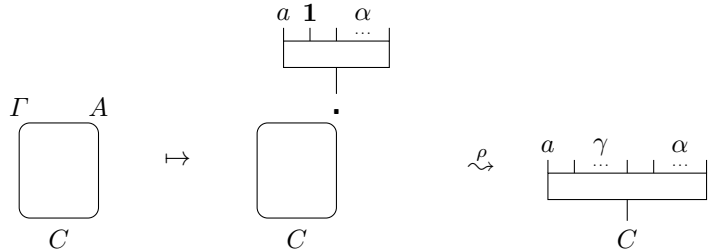
\begin{tikzpicture}
\node (a) at (3em,5em) {$A$};
\node (gamma) at (0em,5em) {$\Gamma$};
\node (c) at (1.5em,0em) {$C$};
\draw [rounded corners] (0em,0.8em) rectangle (3em,4.4em) ;
\node (a) at (12em,5em) {$\centerdot$};
\node (c) at (10.5em,0em) {$C$};
\draw [rounded corners] (9em,0.8em) rectangle (12em,4.4em) ;
\draw (12em,5.5em) -- (12em,6.5em);
\draw (10em,6.5em) -- (14em,6.5em) -- (14em,7.5em) -- (10em,7.5em) -- cycle;
\node (a) at (10em,8.5em) {$a$};
\draw (10em,7.5em) -- (10em,8em);
\node (one) at (11em,8.6em) {$\one$};
\draw (11em,7.5em) -- (11em,8em);
\draw (12em,7.5em) -- (12em,8em);
\draw (14em,7.5em) -- (14em,8em);
\node (dots) at (13em,7.8em) {$\scriptscriptstyle{\ldots}$};
\node (a) at (13em,8.5em) {$\alpha$};
\node (c) at (23em,0em) {$C$};
\draw (c) -- (23em,1.5em);
\draw (20em,1.5em) -- (26em,1.5em) -- (26em,2.5em) -- (20em,2.5em) -- cycle;
\node (a) at (20em,3.5em) {$a$};
\draw (20em,2.5em) -- (20em,3.0em);
\draw (21em,2.5em) -- (21em,3.0em);
\draw (23em,2.5em) -- (23em,3.0em);
\draw (24em,2.5em) -- (24em,3.0em);
\draw (26em,2.5em) -- (26em,3.0em);
\node (dotsa) at (22em,2.8em) {$\scriptscriptstyle{\ldots}$};
\node (dotsb) at (25em,2.8em) {$\scriptscriptstyle{\ldots}$};
\node (gamma) at (22em,3.5em) {$\gamma$};
\node (alpha) at (25em,3.5em) {$\alpha$};
\node (ra) at (17.5em,3em) {$\overset{\rho}{\leadsto}$};
\node at (6em,3em) {$\apsarrow$};
\end{tikzpicture}

Adding the par link for $\downl$ to the above proof net produces to following proof structure, which contracts using the same sequence of contractions $\rho$ as follows.

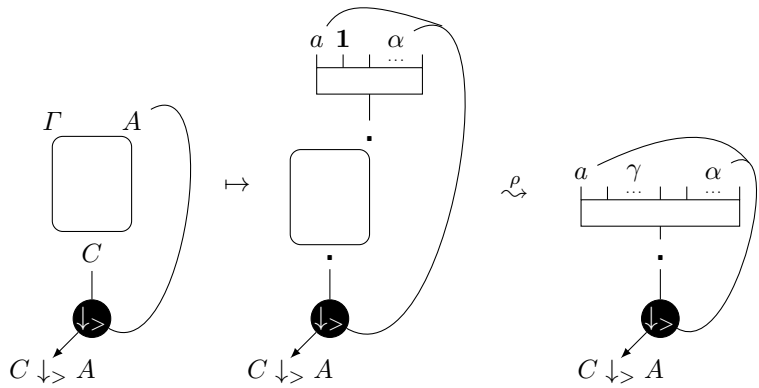
\begin{tikzpicture}
\node (a) at (3em,5.5em) {$A$};
\node (gamma) at (0em,5.5em) {$\Gamma$};
\node (c) at (1.5em,0.5em) {$C$};
\draw [rounded corners] (0em,1.3em) rectangle (3em,4.9em) ;
\node[par] (par) at (1.5em,-2.0em) {};
\node at (1.5em,-2.0em) {\textcolor{white}{$\downl$}};
\draw (par) -- (c);
\node (ca) at (0em,-4.0em) {$C\downl A$};
\draw (par) [in=30,out=-30] to (a);
\path[>=latex,->] (par) edge (0.0em,-3.5em);
\node (a) at (12em,5em) {$\centerdot$};
\node (c) at (10.5em,0.5em) {$\centerdot$};
\draw [rounded corners] (9em,0.8em) rectangle (12em,4.4em) ;
\draw (12em,5.5em) -- (12em,6.5em);
\draw (10em,6.5em) -- (14em,6.5em) -- (14em,7.5em) -- (10em,7.5em) -- cycle;
\node (a) at (10em,8.5em) {$a$};
\draw (10em,7.5em) -- (10em,8em);
\node (one) at (11em,8.6em) {$\one$};
\draw (11em,7.5em) -- (11em,8em);
\draw (12em,7.5em) -- (12em,8em);
\draw (14em,7.5em) -- (14em,8em);
\node (dots) at (13em,7.8em) {$\scriptscriptstyle{\ldots}$};
\node (alpha) at (13em,8.5em) {$\alpha$};
\node[par] (par) at (10.5em,-2.0em) {};
\node at (10.5em,-2.0em) {\textcolor{white}{$\downl$}};
\draw (par) -- (c);
\node (ca) at (9em,-4.0em) {$C\downl A$};
\draw (par) [in=30,out=-30] to (alpha);
\draw (14.75em,9.08em) .. controls (12em,10em) and (11em,10em) .. (a);
\path[>=latex,->] (par) edge (9.0em,-3.5em);
\node (c) at (23em,0.5em) {$\centerdot$};
\draw (23em,1.0em) -- (23em,1.5em);
\draw (20em,1.5em) -- (26em,1.5em) -- (26em,2.5em) -- (20em,2.5em) -- cycle;
\node (a) at (20em,3.5em) {$a$};
\draw (20em,2.5em) -- (20em,3.0em);
\draw (21em,2.5em) -- (21em,3.0em);
\draw (23em,2.5em) -- (23em,3.0em);
\draw (24em,2.5em) -- (24em,3.0em);
\draw (26em,2.5em) -- (26em,3.0em);
\node (dotsa) at (22em,2.8em) {$\scriptscriptstyle{\ldots}$};
\node (dotsb) at (25em,2.8em) {$\scriptscriptstyle{\ldots}$};
\node (gamma) at (22em,3.5em) {$\gamma$};
\node (alpha) at (25em,3.5em) {$\alpha$};
\node[par] (par) at (23.0em,-2.0em) {};
\node at (23.0em,-2.0em) {\textcolor{white}{$\downl$}};
\draw (par) -- (c);
\node (ca) at (21.5em,-4.0em) {$C\downl A$};
\draw (par) [in=30,out=-30] to (alpha);
\draw (26.35em,3.9em) .. controls (25.35em,5.2em) and (23.5em,5.2em) .. (a);
\path[>=latex,->] (par) edge (21.5em,-3.5em);
\node (ra) at (17.5em,3em) {$\overset{\rho}{\leadsto}$};
\node at (7em,3em) {$\apsarrow$};
\end{tikzpicture}

Simply performing the contraction for $\downl$ to the final abstract proof structure produces a comb of $\gamma:C\downl A$ and hence a proof net of $\Gamma\vdash
\gamma:C\downl A$ as required.
\qed

\begin{lemma}\label{lem:complete} Let $\Pi$ be a proof net of $\alpha_1:A_1,\ldots,a_n:A_n \vdash \gamma:C$, that is a proof net with hypotheses $\alpha_1:A_1,\ldots,a_n:A_n$ and
conclusion $C$ and an abstract proof structure contracting to $\gamma$ using contractions $\rho$. There is a natural deduction proof of $\alpha_1:A_1,\ldots,a_n:A_n \vdash \gamma:C$.
\end{lemma}

\paragraph{Proof} We proceed by induction on the number of logical
contractions $l$ in the sequence $\rho$ (this number is equal to the number of par links in the
structure). 

If there are no logical contractions ($l=0$), then there are only structural contractions and our proof net contains only tensor links. We proceed by induction on the
number $t$ of tensor links. If there are no tensor links ($t=0$), we have an
axiom and its abstract proof structure is a comb by definition.

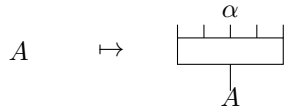
\begin{tikzpicture}
\node (p) at (2em,2em) {$A$};
\node at (10em,0.25em) {$A$};
\draw (8em,1.5em) -- (8em,2.5em) -- (12em,2.5em) -- (12em,1.5em) -- cycle;
\draw (9em,2.5em) -- (9em,3em);
\draw (11em,2.5em) -- (11em,3em);
\node (dots) at (10em,3.5em) {$\alpha$};
\draw (10em,2.5em) -- (10em,3em);
\draw (10em,1.5em) -- (10em,0.5em);
\draw (8em,2.5em) -- (8em,3.0em);
\draw (12em,2.5em)  -- (12em,3.0em);
\node at (5.5em,2em) {$\apsarrow$};
\end{tikzpicture}

This directly gives us the natural deduction proof $\alpha:A \vdash
\alpha:A$. 

If there are tensor links ($t>0$), then either one of the hypotheses or
the conclusion of the proof structure must be the main formula of its
link (this is easy to see since if none of the leaves is the main
formula of its link, then the proof structure contains only
introduction rules for $\bullet$ and $\prodk$ and therefore the conclusion is the main formula of
its link). Suppose a proof net has a leaf which is the main formula of its link and suppose this formula is $A \downl C$ (the cases of other formulas being main formulas, and of a conclusion of the proof net being
the main formula are similar). Then, since all tensor trees contract to combs, we can apply the induction hypothesis to the two structures obtained by removing the tensor
link and obtain proofs $\pi_1$ of $\Gamma \vdash
a+\one+\alpha:A$ and $\pi_2$ of $\Delta,
a+\gamma+\alpha:C \vdash \delta:D$ (technically, we have a proof with hypothesis $\gamma':C$ and use substitution of the proof with conclusion $a+\gamma+\alpha:C$ shown below). We can combine these proofs as follows.

\[
\infer*[\pi2]{\delta:D}{\infer[\downl E]{a+\gamma+\alpha:C}{\infer*[\pi1]{a+\one+\alpha:A}{\Gamma} & \gamma:A\downl C} & \Delta}
\]

\editout{
\begin{tikzpicture}
\draw [rounded corners] (-2em,5.4em) rectangle (1em,8.4em) ;
\draw [rounded corners] (1.5em,-0.6em) rectangle (4.5em,-3.6em) ;
\node (ab) at (3em,0.0em) {$C$};
\node (a) at (0,4.8em) {$A_{\rule{0pt}{1.2ex}}$};
\node (b) at (6em,4.8em) {$\smash{A\downl C}\rule{0pt}{1.3ex}$};
\node[tns] (c) at (3em,2.868em) {};
\node (clab) at (3em,2.868em) {$\timesl$};
\draw (c) -- (ab);
\draw (c) -- (a);
\draw (c) -- (b);
\node (d) at (3em,-4.2em) {$D$};
\draw [rounded corners] (10em,5.4em) rectangle (13em,8.4em) ;
\draw [rounded corners] (13.5em,-0.6em) rectangle (16.5em,-3.6em) ;
\node (ab) at (15em,0.0em) {$\centerdot$};
\node (a) at (12em,4.8em) {$\centerdot$};
\node (b) at (18em,4.8em) {$\centerdot$};
\node[tns] (c) at (15em,2.868em) {};
\node (clab) at (15em,2.868em) {$\timesl$};
\draw (c) -- (ab);
\draw (c) -- (a);
\draw (c) -- (b);
\draw (18em,5.3em) -- (18em,6em);
\draw (17em,6em) -- (19em,6em) -- (19em,7em) -- (17em,7em) -- cycle;
\draw (17em,7em) -- (17em,7.5em);
\draw (19em,7em) -- (19em,7.5em);
\node (dots) at (18em,7.2em) {$\scriptscriptstyle{\ldots}$};
\node (gamma) at (18em,7.8em) {$\gamma$};
\node (d) at (15em,-4.2em) {$D$};
\node (d) at (24em,0em) {$D$};
\draw (24em,0.5em) -- (24em,1.0em);
\draw (23em,1em) -- (25em,1em) -- (25em,2em) -- (23em,2em) -- cycle;
\draw (23em,2em) -- (23em,2.5em);
\draw (25em,2em) -- (25em,2.5em);
\node (dots) at (24em,2.2em) {$\scriptscriptstyle{\ldots}$};
\node (gamma) at (24em,2.8em) {$\delta$};
\node (ra) at (20.5em,2em) {$\overset{\rho}{\leadsto}$};
\node at (8em,2em) {$\apsarrow$};
\end{tikzpicture}
}

If the sequence $\rho$ has logical contractions ($l>0$), we look at the last such contraction and proceed by case analysis. If the last contraction is the $\downl$ contraction, our proof net
and contraction sequence look as follows. 

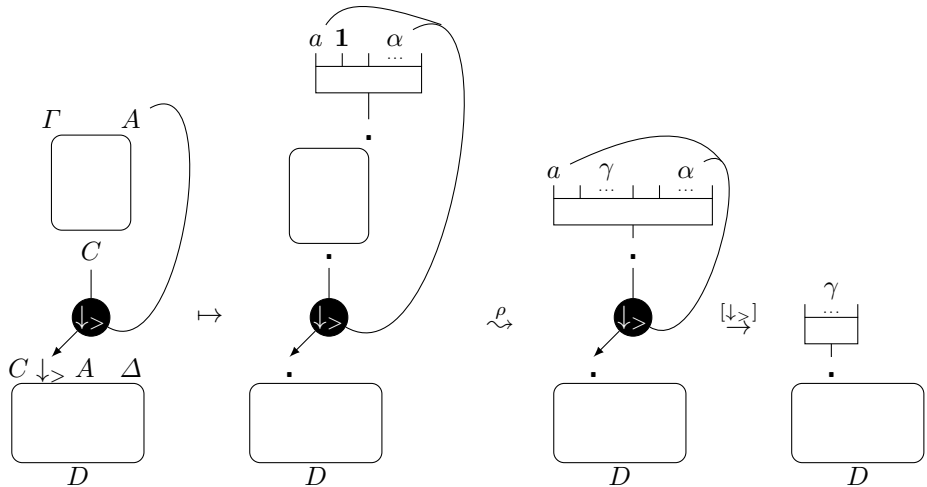
\hspace*{-2em}\begin{tikzpicture}
\node (a) at (3em,5.5em) {$A$};
\node (gamma) at (0em,5.5em) {$\Gamma$};
\node (c) at (1.5em,0.5em) {$C$};
\draw [rounded corners] (0em,1.3em) rectangle (3em,4.9em) ;
\node[par] (par) at (1.5em,-2.0em) {};
\node at (1.5em,-2.0em) {\textcolor{white}{$\downl$}};
\draw (par) -- (c);
\node (ca) at (0em,-4.0em) {$C\downl A$};
\draw (par) [in=30,out=-30] to (a);
\path[>=latex,->] (par) edge (0.0em,-3.5em);
\draw [rounded corners] (-1.5em,-4.5em) rectangle (3.5em,-7.5em) ;
\node (delta) at (3em,-3.9em) {$\Delta$};
\node (d) at (1.0em,-8.0em) {$D$};
\node (a) at (12em,5em) {$\centerdot$};
\node (c) at (10.5em,0.5em) {$\centerdot$};
\draw [rounded corners] (9em,0.8em) rectangle (12em,4.4em) ;
\draw (12em,5.5em) -- (12em,6.5em);
\draw (10em,6.5em) -- (14em,6.5em) -- (14em,7.5em) -- (10em,7.5em) -- cycle;
\node (a) at (10em,8.5em) {$a$};
\draw (10em,7.5em) -- (10em,8em);
\node (one) at (11em,8.6em) {$\one$};
\draw (11em,7.5em) -- (11em,8em);
\draw (12em,7.5em) -- (12em,8em);
\draw (14em,7.5em) -- (14em,8em);
\node (dots) at (13em,7.8em) {$\scriptscriptstyle{\ldots}$};
\node (alpha) at (13em,8.5em) {$\alpha$};
\draw [rounded corners] (7.5em,-4.5em) rectangle (12.5em,-7.5em) ;
\node[par] (par) at (10.5em,-2.0em) {};
\node at (10.5em,-2.0em) {\textcolor{white}{$\downl$}};
\draw (par) -- (c);
\node (ca) at (9em,-4.0em) {$\centerdot$};
\draw (par) [in=30,out=-30] to (alpha);
\draw (14.75em,9.08em) .. controls (12em,10em) and (11em,10em) .. (a);
\path[>=latex,->] (par) edge (9.0em,-3.5em);
\node (d) at (10.0em,-8.0em) {$D$};
\node (c) at (22em,0.5em) {$\centerdot$};
\draw (22em,1.0em) -- (22em,1.5em);
\draw (19em,1.5em) -- (25em,1.5em) -- (25em,2.5em) -- (19em,2.5em) -- cycle;
\node (a) at (19em,3.5em) {$a$};
\draw (19em,2.5em) -- (19em,3.0em);
\draw (20em,2.5em) -- (20em,3.0em);
\draw (22em,2.5em) -- (22em,3.0em);
\draw (23em,2.5em) -- (23em,3.0em);
\draw (25em,2.5em) -- (25em,3.0em);
\node (dotsa) at (21em,2.8em) {$\scriptscriptstyle{\ldots}$};
\node (dotsb) at (24em,2.8em) {$\scriptscriptstyle{\ldots}$};
\node (gamma) at (21em,3.5em) {$\gamma$};
\node (alpha) at (24em,3.5em) {$\alpha$};
\draw [rounded corners] (19.0em,-4.5em) rectangle (24.0em,-7.5em) ;
\node[par] (par) at (22.0em,-2.0em) {};
\node at (22.0em,-2.0em) {\textcolor{white}{$\downl$}};
\draw (par) -- (c);
\node (ca) at (20.5em,-4.0em) {$\centerdot$};
\draw (par) [in=30,out=-30] to (alpha);
\draw (25.35em,3.9em) .. controls (24.35em,5.2em) and (22.5em,5.2em) .. (a);
\path[>=latex,->] (par) edge (20.5em,-3.5em);
\node (d) at (21.5em,-8.0em) {$D$};
\node (ra) at (17.0em,-2em) {$\overset{\rho}{\leadsto}$};
\node at (6em,-2em) {$\apsarrow$};
\node (ra) at (26.0em,-2em) {$\overset{[\downl]}{\rightarrow}$};
\draw [rounded corners] (28.0em,-4.5em) rectangle (33.0em,-7.5em) ;
\node (d) at (30.5em,-8.0em) {$D$};
\node (ca) at (29.5em,-4.0em) {$\centerdot$};
\draw (29.5em,-3.5em) -- (29.5em,-3em);
\draw (28.5em,-3em) -- (30.5em,-3em) -- (30.5em,-2em) -- (28.5em,-2em) -- cycle;
\draw (28.5em,-2em) -- (28.5em,-1.5em);
\draw (30.5em,-2em) -- (30.5em,-1.5em);
\node (dotsa) at (29.5em,-1.8em) {$\scriptscriptstyle{\ldots}$};
\node (gamma) at (29.5em,-1em) {$\gamma$};
\end{tikzpicture}

The initial proof structure is shown above of the left and its corresponding abstract proof structure to its immediate right (note that vertex $A$ has been replaced by
$a\dplus\one\dplus\alpha$, since it is an auxiliary input, corresponding to a withdrawn hypothesis in the natural deduction proof). The reduction sequence is of the form
$\rho$, followed by the $\downl$ contraction, possibly followed by a number of structural contractions (not displayed in the figure above).

When we remove the par link from the figure above, we are in the following situation. All contractions from $\rho$ are either fully in the abstract proof structure shown
below at the top of the picture or fully in the abstract proof structure shown below at the bottom of the picture, so $\rho$ splits naturally in $\rho_1$ and $\rho_2$. 

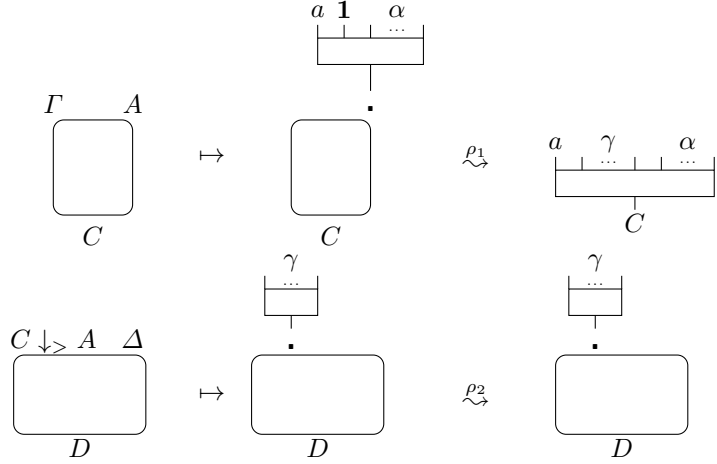
\begin{tikzpicture}
\node (a) at (3em,5em) {$A$};
\node (gamma) at (0em,5em) {$\Gamma$};
\node (c) at (1.5em,0em) {$C$};
\draw [rounded corners] (0em,0.8em) rectangle (3em,4.4em) ;
\node (ca) at (0em,-4.0em) {$C\downl A$};
\draw [rounded corners] (-1.5em,-4.5em) rectangle (3.5em,-7.5em) ;
\node (delta) at (3em,-3.9em) {$\Delta$};
\node (d) at (1.0em,-8.0em) {$D$};
\node (a) at (12em,5em) {$\centerdot$};
\node (c) at (10.5em,0.0em) {$C$};
\draw [rounded corners] (9em,0.8em) rectangle (12em,4.4em) ;
\draw (12em,5.5em) -- (12em,6.5em);
\draw (10em,6.5em) -- (14em,6.5em) -- (14em,7.5em) -- (10em,7.5em) -- cycle;
\node (a) at (10em,8.5em) {$a$};
\draw (10em,7.5em) -- (10em,8em);
\node (one) at (11em,8.6em) {$\one$};
\draw (11em,7.5em) -- (11em,8em);
\draw (12em,7.5em) -- (12em,8em);
\draw (14em,7.5em) -- (14em,8em);
\node (dots) at (13em,7.8em) {$\scriptscriptstyle{\ldots}$};
\node (alpha) at (13em,8.5em) {$\alpha$};
\draw [rounded corners] (7.5em,-4.5em) rectangle (12.5em,-7.5em) ;
\node (ca) at (9em,-4.0em) {$\centerdot$};
\node (d) at (10.0em,-8.0em) {$D$};
\node (c) at (22em,0.5em) {$C$};
\draw (22em,1.0em) -- (22em,1.5em);
\draw (19em,1.5em) -- (25em,1.5em) -- (25em,2.5em) -- (19em,2.5em) -- cycle;
\node (a) at (19em,3.5em) {$a$};
\draw (19em,2.5em) -- (19em,3.0em);
\draw (20em,2.5em) -- (20em,3.0em);
\draw (22em,2.5em) -- (22em,3.0em);
\draw (23em,2.5em) -- (23em,3.0em);
\draw (25em,2.5em) -- (25em,3.0em);
\node (dotsa) at (21em,2.8em) {$\scriptscriptstyle{\ldots}$};
\node (dotsb) at (24em,2.8em) {$\scriptscriptstyle{\ldots}$};
\node (gamma) at (21em,3.5em) {$\gamma$};
\node (alpha) at (24em,3.5em) {$\alpha$};
\draw [rounded corners] (19.0em,-4.5em) rectangle (24.0em,-7.5em) ;
\node (ca) at (20.5em,-4.0em) {$\centerdot$};
\node (d) at (21.5em,-8.0em) {$D$};
\node (ra) at (16.0em,3em) {$\overset{\rho_1}{\leadsto}$};
\node (ra) at (16.0em,-6em) {$\overset{\rho_2}{\leadsto}$};
\node at (6em,-6em) {$\apsarrow$};
\node at (6em,3em) {$\apsarrow$};
\draw (20.5em,-3.5em) -- (20.5em,-3em);
\draw (19.5em,-3em) -- (21.5em,-3em) -- (21.5em,-2em) -- (19.5em,-2em) -- cycle;
\draw (19.5em,-2em) -- (19.5em,-1.5em);
\draw (21.5em,-2em) -- (21.5em,-1.5em);
\node (dotsa) at (20.5em,-1.8em) {$\scriptscriptstyle{\ldots}$};
\node (gamma) at (20.5em,-1em) {$\gamma$};
\draw (9.0em,-3.5em) -- (9.0em,-3em);
\draw (8.0em,-3em) -- (10.0em,-3em) -- (10.0em,-2em) -- (8.0em,-2em) -- cycle;
\draw (8.0em,-2em) -- (8.0em,-1.5em);
\draw (10.0em,-2em) -- (10.0em,-1.5em);
\node (dotsa) at (9.0em,-1.8em) {$\scriptscriptstyle{\ldots}$};
\node (gamma) at (9.0em,-1em) {$\gamma$};
\end{tikzpicture}

We need to show that $\Gamma,\Delta \vdash \delta:D$ (where $\delta:D$ is the comb). Since we have two proof nets with strictly shorter sequences of contractions, we can
apply the induction hypothesis for  proofs $\pi_1$ of $a\dplus\one\dplus\alpha:A,\Gamma\vdash
a\dplus\gamma\dplus\alpha:C$ and $\pi_2$ of $\gamma:C\downl A,\Delta\vdash \delta:D$. We can combine these two proofs into a proof of $\Gamma,\Delta \vdash \delta:D$ as follows.
\[
\infer*[\pi_2]{\delta:D}{\infer[\downl I_i]{\gamma:C\downl A}{\infer*[\pi_1]{a\dplus\gamma\dplus\alpha:C}{[a\dplus\one\dplus\alpha:A]_i & \Gamma\quad}} & \Delta\quad}
\]

The other cases are similar and easily verified.\qed

\begin{theorem}
A proof structure is a \emph{proof net} iff its abstract proof structure contracts to a comb. 
\end{theorem}

\paragraph{Proof} Immediate from Lemma~\ref{lem:sound} and Lemma~\ref{lem:complete}. \qed

\section{Extension to Other Connectives}
\label{sec:extension}

One of the benefits of the current calculus is that it extends easily to other connectives, such as the unary/bracket connectives of \citeasnoun[Chapter~5]{morrill2010}
(although incorporating the treatment of parasitic gapping of Section~5.5 would require a considerable complication of the proof theory). 

The synthetic connectives of \citeasnoun{mvf11displacement} require us to extend our methodology somewhat: as currently formulated the proof net calculus produces a single comb for each proof net. When
adding the synthetic connectives, we can introduce a separation marker in a way which is only partially specified by the premiss of the rule. For example, the denotation
of (leftmost) split $\splitc A$, shown below, is the set of strings obtained by inserting a separator symbol at any place before other separator symbols (if any), and
therefore the introduction rule for this connective doesn't produce a unique string term.
\begin{align}
| \splitc A | &=_{\mathit{def}} \{ a+\one+\alpha \,|\, 
a+\alpha \in | A | \} \\
| \bridge A | &=_{\mathit{def}} \{ a+\alpha \,|\,  a+\one+\alpha \in | A | \} 
\end{align}

This moves us to a system where a tensor tree contracts to a set of combs (or, alternatively, a partially specified comb). Apart from this, it is not hard to add links
and contractions for the synthetic connectives. For example, the contraction for $\splitc$ can be obtained from the contraction for $\upk$ by removing the links to the
auxiliary hypothesis: instead of replacing the auxiliary hypothesis by $\one$ (which defines the position of the insertion point uniquely), there will be multiple,
non-confluent ways to matching the contraction and to insert the separator symbol. For lack of space, we will not develop these ideas further here.

\section{Conclusion}

We have presented a proof net calculus for the Displacement calculus and shown its correctness. This is the first proof net calculus which models the Displacement
calculus directly and not by some sort of translation into another formalism. The proof net calculus opens up new possibilities for parsing and proof search with the
Displacement calculus.

\bibliographystyle{agsm}
\bibliography{moot}

\end{document}